\documentclass[msom,nonblindrev]{informs3noheader}

\OneAndAHalfSpacedXI

\usepackage{changepage}
\usepackage{marvosym}
\usepackage{subcaption}
\usepackage{pgfplots}
\usepackage{tikz}
\usetikzlibrary{decorations.pathreplacing, calligraphy}
\usetikzlibrary{tikzmark, positioning, fit, shapes.misc}
\usetikzlibrary{calc}
\usepackage{graphicx}
\usetikzlibrary{arrows,shapes}
\usepackage{array}
\usepackage{tikz-qtree}
\usepackage{booktabs} \usepackage[ruled]{algorithm2e} \usepackage{listings}

\SetAlFnt{\small}
\SetAlCapFnt{\small}
\SetAlCapNameFnt{\small}
\SetAlCapHSkip{0pt}
\IncMargin{-\parindent}
\usepackage{hyperref}
\usepackage{soul}
\usepackage{multirow}
\usepackage{framed}
\usepackage{arydshln}
\def\bold#1{{\sffamily \bfseries#1}}

\newcommand{\EE}{{\mathbb E}} 

\newcommand{\PP}{{\mathbb P}} \newcommand{\QQ}{{\mathbb Q}}

\newcommand{\RR}{{\mathbb R}}

\def\History{\mathcal{H}}

\def\proba{\mathbb{P}}

\def\history{h}
\def\labl{y} 
\def\Mapping{\mu}
\def\Valuation{V}
\def\Reward{R}

\usepackage{natbib}
 \bibpunct[, ]{(}{)}{,}{a}{}{,}\def\bibfont{\small}\def\BIBand{and}

\TheoremsNumberedThrough     \ECRepeatTheorems

\EquationsNumberedThrough    

\MANUSCRIPTNO{}

\begin{document}

\RUNAUTHOR{\bold{Bompaire, Désir, Heymann}}

\RUNTITLE{\sffamily Robust label attribution for real-time bidding}

\TITLE{\Large \bold{Fixed point label attribution for real-time bidding}}

\ARTICLEAUTHORS{\AUTHOR{\sffamily Martin Bompaire}
\AFF{Criteo AI Lab,
\EMAIL{\texttt{m.bompaire@criteo.com}}} 
\AUTHOR{\sffamily Antoine Désir}
\AFF{Technology and Operations Management, INSEAD, \EMAIL{\texttt{antoine.desir@insead.edu}}}
\AUTHOR{\sffamily Benjamin Heymann}
\AFF{Criteo AI Lab,
\EMAIL{\texttt{b.heymann@criteo.com}}} 
\vspace{0.2cm}
\AUTHOR{\small This version: {\sffamily \today}}
} 

\ABSTRACT{\textbf{Problem definition:} Most of the display advertising inventory is sold through real-time auctions. The participants of these auctions are typically bidders (Google, Criteo, RTB House, Trade Desk for instance) who participate on behalf of advertisers. In order to  estimate the value of each display opportunity, they usually train advanced machine learning algorithms using historical data. In the labeled training set, the inputs  are vectors of features representing each display opportunity and the labels are the generated rewards. In practice, the rewards are given by the advertiser and are tied to whether or not a particular user converts. Consequently, the rewards are aggregated at the user level and never observed at the display level. A fundamental task that has, to the best of our knowledge, been overlooked is to account for this mismatch and split, or attribute, the rewards at the right granularity level before training a learning algorithm. We call this the label attribution problem. 

\noindent \textbf{Methodology/results:}  In this paper, we develop an approach to the label attribution problem, which is both theoretically justified and practical. In particular, we develop a fixed point algorithm that allows for large scale implementation and showcase our solution using a large scale publicly available dataset from Criteo, a large Demand Side Platform. We dub our approach the Fixed Point Label Attribution (\textsf{FiPLA}) Algorithm.

\noindent \textbf{Managerial implications:} There is often a hidden leap of faith when transforming the advertiser's signal into  display labelling. DSP providers should be careful when building their machine learning pipeline and carefully solve the label attribution step. 
}

\KEYWORDS{Online advertising, attribution,  autobidding}

\maketitle

\section{Introduction} \label{sec:introduction}

Digital advertising has been growing continuously since its inception and sustains a large part of the internet as we know it. It is indeed the main source of revenue for several tech giants (Google and Facebook in particular), and has attracted billions of dollars from advertisers over the last two decades. In 2019, the global digital ad spending exceeded  300 billion USD \citep{emarketer}. What makes digital marketing so attractive is that  marketers can harness the power of data to make informed budget allocation decisions, and show the right banner to the right customer at the right time.

In this paper, we take the point of view of a \emph{bidder} (Google, Criteo, RTB House, Trade Desk for instance) who participates in a real-time bidding exchange on behalf of an advertiser~\citep{balseiro2017optimal}. Usually, such a bidder is rewarded when an ad is won, clicked on and followed in a short period of time by a conversion, typically a sale or any kind of action on the advertiser website such as creating an account or putting an item in a basket. Whenever the bidder receives a display opportunity, he has to estimate its potential reward and then, whatever the auction mechanism, calibrate his bid accordingly. The prediction of this potential reward is usually the result of a supervised learning algorithm that requires \emph{labeled} training samples, where the inputs are vectors of features $\boldsymbol{x}$, which represent the display opportunity (these features encode contextual and user information) and the labels are the generated rewards. Table~\ref{tab:ex_1} shows a typical dataset.
\begin{table}[h]\centering
\caption{A typical dataset used by the bidder to estimate the potential reward of future display opportunities.}\label{tab:ex_1}
\setlength\extrarowheight{2pt} 
\begin{tabular}{ccccccc}
  \multicolumn{6}{c}{Features ($\boldsymbol{x}$)} & Reward  \\
 \cmidrule[1pt](lr){1-6} \cmidrule[1pt](lr){7-7} 
 User ID & Timestamp & Clicked & Display Env. & User Funnel & $\quad\dots\quad$ & Conversion \\
 \cmidrule[1pt](lr){1-6}  \cmidrule[1pt](lr){7-7} 
 1 & 10 & 0 & Desktop &  Not exposed & & 0\\
 1 & 11 & 1 & Desktop &  Exposed & & 1 \\
 1 & 12 & 0 & Desktop &  Engaged & & 0 \\
 \noalign{\vskip 0.5ex}\hdashline\noalign{\vskip 0.5ex}
 2 & 20 & 1 & Mobile &  Not exposed & & 0\\
 2 & 21 & 1 & Mobile &  Engaged & & 1 \\
 2 & 22 & 0 & Desktop &  Engaged & & 0 \\
 2 & 23 & 0 & Desktop &  Engaged & & 0\\
 \noalign{\vskip 0.5ex}\hdashline\noalign{\vskip 0.5ex}
 3 & 30 & 0 & Mobile & Not exposed & & 0 \\
 \cmidrule[1pt](lr){1-7}
\end{tabular}
\end{table}

Using a machine learning (ML) terminology, the reward corresponds to the label and this is in fact a typical supervised learning problem, where any ML algorithm can be used to predict the reward as a function of the display $\boldsymbol{x}$. 

This seemingly straightforward and widely adopted approach unfortunately fails to recognize a crucial aspect of the data: a conversion is the result of a sequence of interactions with a user. Indeed, the rewards, which are decided by the advertiser, are in practice tied to a particular user and given not for a single display but for a sequence of interactions. An important step, which, to the best of our knowledge, has been overlooked is to account for this mismatch and split, or attribute the rewards at the display granularity. We dub this step the \textbf{label attribution}. Table~\ref{tab:snapshot} illustrates this step.
\begin{table}[h]\centering
\caption{Label attribution step: the reward is spread among the user's displays instead of being arbitrarily assigned to one of them.}\label{tab:snapshot}
\setlength\extrarowheight{2pt} 
\begin{tabular}{cccccccc}
User &  \multicolumn{4}{c}{Features ($\boldsymbol{x}$)} & Reward & Valuation ($y$)  \\
 \cmidrule[1pt] (lr) {1-1} \cmidrule[1pt](lr){2-5} \cmidrule[1pt](lr){6-6}  \cmidrule[1pt](lr){7-7} 
  ID & Timestamp & Clicked &  Display Env. & $\quad\dots\quad$ & Conversion  & Label  \\
 \cmidrule[1pt](lr){1-1} \cmidrule[1pt](lr){2-5}  \cmidrule[1pt](lr){6-6} \cmidrule[1pt](lr){7-7} 
 \multirow{3}{*}{1} & 10 & 0 & Desktop & & \multirow{3}{*}{\tikzmark{x1}1} & \tikzmark{label1}0.1 \\
  & 11 & 1 & Desktop & & & \tikzmark{label2}0.7 \\
  & 12 & 0 & Desktop & & & \tikzmark{label3}0.2\\
  \noalign{\vskip 0.5ex}\hdashline\noalign{\vskip 0.5ex}
 \multirow{4}{*}{2} & 20 & 1 & Mobile & & \multirow{4}{*}{\tikzmark{x2}1} & \tikzmark{label4}0.5 \\
  & 21 & 1 & Mobile & & &\tikzmark{label5}0.3  \\
  & 22 & 0 & Desktop & & &\tikzmark{label6}0.1  \\
  & 23 & 0 & Desktop & & &\tikzmark{label7}0.1 \\
  \noalign{\vskip 0.5ex}\hdashline\noalign{\vskip 0.5ex}
3 & 30 & 0 & Mobile & & \tikzmark{x3}0 & \tikzmark{label8}0 \\
 \cmidrule[1pt](lr){1-5} \cmidrule[1pt](lr){6-7}
\end{tabular}
\begin{tikzpicture}[overlay,remember picture, shorten >=-3pt]
\node[rectangle,draw = blue!50, fill = blue!50,opacity=.2,text opacity=1,inner sep=6pt] at ($(pic cs:x1) + (0.1,0.1)$) (x) {};
\draw[->] (x) to[out=0,in=180] ($(pic cs:label1) + (-0.2,0.1) $);
\draw[->] (x) to[out=0,in=180] ($(pic cs:label2) + (-0.2,0.1) $) ;
\draw[->] (x) to[out=0,in=180] ($(pic cs:label3) + (-0.2,0.1) $) ;
\node[rectangle,draw = blue!50, fill = blue!50,opacity=.2,text opacity=1,inner sep=6pt] at ($(pic cs:x2) + (0.1,0.1)$) (y) {};
\draw[->] (y) to[out=0,in=180] ($(pic cs:label4) + (-0.2,0.1) $);
\draw[->] (y) to[out=0,in=180] ($(pic cs:label5) + (-0.2,0.1) $) ;
\draw[->] (y) to[out=0,in=180] ($(pic cs:label6) + (-0.2,0.1) $) ;
\draw[->] (y) to[out=0,in=180] ($(pic cs:label7) + (-0.2,0.1) $) ;
\node[rectangle,draw = blue!50, fill = blue!50,opacity=.2,text opacity=1,inner sep=6pt] at ($(pic cs:x3) + (0.1,0.1)$) (z) {};
\draw[->] (z) to[out=0,in=180] ($(pic cs:label8) + (-0.2,0.1) $) ;
\end{tikzpicture}
\end{table}

Existing approaches \citep{Chapelle2014a} omit this important step of converting the raw data given by the advertiser into a dataset that can be fed to a ML pipeline and attribute, somewhat heuristically, the reward given by the advertiser entirely to the last display. We refer to this approach as the \emph{last touch} approach. Among the technical challenges a bidder is facing, recognizing this step and choosing a label attribution mechanism is often an understated problem. 

\medskip 

\emph{In this paper, we formalize this missing step and propose a principled and scalable solution to predict the value of a given display opportunity that accounts for the fact that the reward given by the advertiser are aggregated at the user level.}

\subsection{Connection to attribution in online advertising}

This problem is related to the well-studied question of attribution in online advertising, i.e., assigning conversions credits to individual marketing interactions. It is usually posed from the point of view of the advertiser who needs to decide how to split the credit between different advertising channels and is a fundamental question that informs the media mix optimization or the understanding of a customer's journey. It has recently been identified as a top research priority by the Marketing Science Institute \citep{Lemon2016}. 

Similarly, in deciding how to split the reward given by the advertiser for sequences of display opportunities, the bidder also needs to perform his own attribution. To differentiate between these two levels of attribution, we refer to the bidder's task as the \emph{label attribution} problem. In fact, the reward received by the bidder is the result of the advertiser's attribution. However, for our purposes, we assume that the attribution mechanism used by the advertiser, which impacts the reward of the bidder, is fixed and exogenous for the bidder. This leader-follower assumption is mild, since the advertiser attribution rule is almost always fixed in practice.

\subsection{Motivating example} \label{sec:motivating_example}

We next present an example to illustrate the limitations of ignoring the label attribution step and motivate the need for a more principled approach. Just like the last touch attribution used by advertisers, using a last touch approach for the label attribution fails to recognize the potential effect of early displays moving a customer up the conversion funnel and therefore fails to acknowledge their contribution into generating the reward. This leads to undervaluing early display opportunities and overvaluing later ones.

To make this even more concrete and build some intuition, assume for simplicity that there are only two types of display opportunities. Displays A influence users' behavior and increase their conversion probability. On the other hand, displays B do not influence the conversion probability at all but artificially always appear before a conversion, as is common for search ads~\citep{blake2015consumer}. In such situation, the last touch label attribution wrongly allocates all the reward to displays B.  Over time, the bidder will undervalue displays A and overvalue displays B.  This in turn will lead to fewer conversions and therefore also lower the reward attributed to the bidder. This example highlights that the label attribution is a crucial step to align the bidder strategy with the given reward. It showcases the limitations of the current last touch approach and motivates the development of alternate  methods of label attribution. We revisit this example in our numerical experiments in Section~\ref{sec:numerics_synthetic} and \ref{sec:numerics_revenue}.

We end this example by noting that the motivation for using a last touch label attribution rule in practice seems to be driven by the misconception that if the advertiser uses a last touch attribution rule, which is a prevalent heuristic~\citep{Ji2017,diemert2017attribution}, then the bidder should in turn use a last touch label attribution rule. Since our example does not depend on the particular advertiser attribution rule, it demonstrates that this belief is wrong.

\subsection{Contribution}

\paragraph{The label attribution problem.} In order to accurately bid in any auction, a bidder needs to learn a valuation function $V$ to evaluate each display opportunity. The main challenge is that rewards are given at the user level, i.e., for sequences of displays or timelines, and cannot be used directly as labels. Consequently, we introduce in this paper the label attribution problem, which consists of finding a mapping $\Mapping$ that splits the reward for every user into labels for each individual display opportunity. These labels can then be used to learn the valuation function $V$ using any  appropriate ML algorithm. Our first contribution is to identify and formalize this label attribution problem which has not, to the best of our knowledge, been studied before. In particular, existing approaches implicitly assume that the label attribution step is done using a last touch method. 

\paragraph{A practical algorithm.} A natural way to value a display is through its marginal contribution, i.e., the lift it provides to an existing sequence of displays. We call such a valuation \emph{additive} as it decomposes the reward additively over the sequence of displays and next propose a practical algorithm to learn such an additive valuation. In particular, we argue that the label attribution needed to learn such valuation must satisfy a fixed point equation. We leverage this relation to propose an iterative algorithm that simultaneously learns the label attribution and valuation. We refer to our algorithm as the Fixed Point Label Attribution (\textsf{FiPLA}) Algorithm. Our approach is very practical as it can easily be embedded into an existing ML pipeline by adding a fixed point feedback loop. 

More precisely, our approach is iterative and alternates between an update step and a predict step. In each iteration $k$, we maintain two mappings: a label attribution mechanism $\Mapping^{(k)}$, which is used to split the rewards, and a valuation mechanism $\Valuation^{(k)}$, which transforms any vector of features into a predicted reward. In the first step of each iteration, we update the mapping $\Mapping^{(k)}$ using the fixed point relation and the learned mapping $\Valuation^{(k)}$. We then, in a second step, split the rewards from the raw data using the label attribution $\Mapping^{(k)}$ and learn a new mapping $\Valuation^{(k+1)}$ using any ML algorithm. Figure~\ref{fig:paradigm-change} summarizes our framework.

\begin{figure}[t]
\begin{center}
    \begin{tikzpicture}
        
\node[text width=2cm,align=center] at (-1.5,0) {Traditional\\approach}; 

\node at (1,0) {\sf Rewards};
\draw[->, very thick] (2,0)--(5,0);
\node at (3.5,-0.2) {\tiny \sf label attribution};
\node at (3.5,0.3) {$\Mapping \sim \sf last\;touch$};
\node at (6.5,0) {\sf Training set};
\draw[->, very thick] (8,0)--(11,0);
\node at (9.5,0.3) {\sf learn with ML};
\node at (9.5,-0.2) {\tiny \sf prediction};
\node at (12.5,0) {$\Valuation : \boldsymbol{x} \mapsto \hat{\labl}$};

\node[text width=2cm,align=center] at (-1.5,-4) {Proposed\\approach}; 

\node at (1,-3) {\sf Rewards};
\draw[->, very thick] (2,-3)--(5,-3);
\node at (3.5,-3.2) {\tiny \sf label attribution};
\node at (3.5,-2.7) {$\Mapping^{(k)}$};
\node at (6.5,-3) {\sf Training set};
\draw[->, very thick] (8,-3)--(11,-3);
\node at (9.5,-2.7) {\sf learn with ML};
\node at (9.5,-3.2) {\tiny \sf prediction};
\node at (12.5,-3) {$\Valuation^{(k)} : \boldsymbol{x} \mapsto \hat{\labl}$};
\draw[->, very thick] (12.5,-3.5) -- (12.5,-5) -- (1,-5) -- (1,-3.5);
\node[align=center] at (7,-4.7) {\sf fixed point equation};
\node[align=center] at (7,-5.2) {\tiny \sf feedback loop};

       \end{tikzpicture}
    \caption{The traditional label attribution approach uses the last touch heuristic. By contrast, our proposed approach is an iterative algorithm that alternates between updating the attribution mapping $\Mapping^{(k)}$ and the prediction function $\Valuation^{(k)}$. The update of the mapping $\Mapping^{(k)}$ relies on a fixed point relation. Our framework can be implemented by adding a feedback loop to an existing ML pipeline.}\label{fig:paradigm-change}
 \end{center}
\end{figure}

\paragraph{Numerical experiments.} We illustrate the benefits our approach through three different numerical experiments.\footnote{All our codes are available at \url{https://github.com/criteo-research/robust-label-attribution}} 
\begin{enumerate}
    \item First, we explore what the \textsf{FiPLA} Algorithm learns in simple synthetic settings. We showcase that our approach can correctly capture critical aspects of typical user behaviors such as a decreasing marginal return effect of advertising. More importantly, we highlight the limitations of the last touch approach.
    \item We then test our approach on a large scale dataset. This real dataset exhibit a large feature space and we discuss how the \textsf{FiPLA} can be implemented in such a setting. We compare the output of the \textsf{FiPLA} Algorithm to the typical last touch approach both qualitatively and using an independently proposed performance metrics.
    \item Finally, we simulate a real bidding environment and measure the benefits of using our approach on the bidder's revenue. In our simulated settings, the profit associated with using the \textsf{FiPLA} Algorithm is more than 3.5 times higher than using a last touch approach, even when we allow for shading. This shows that the potential benefits of carefully addressing the label attribution step can be significant.
\end{enumerate}

\paragraph{A principled approach.} We give some theoretical underpinning for our approach and mathematically formalize the relation between the fixed point equation and the additivity property for a slightly stylized model. In particular, under some assumptions, we show that the additivity property defines a unique label attribution mechanism and that the sequence of mapping $V^{(k)}$ converges to the additive valuation. We further show that this intuitive label attribution enjoys several interesting properties. First, we show that bidding according to this additive valuation is myopic optimal, i.e., optimal if there are no other future opportunities. Moreover, we show that our proposed approach is distributionally robust, i.e., it does not depend on how many times each set of features has been observed. This is very useful from a learning perspective.

\paragraph{Agenda.}
 The paper is organized as follows. We begin with a review of the literature in Section~\ref{sec:related-work}. We then introduce the notations that allow us to properly frame the label attribution problem in Section~\ref{sec:model}. We present our proposed learning framework in Section~\ref{sec:additivity}. In Section~\ref{sec:numerics_synthetic}, we illustrate our method on a series of synthetic experiments. We then test our framework on a large publicly available dataset in Section~\ref{sec:numerics_criteo}. In Section~\ref{sec:numerics_revenue}, we evaluate the impact on revenue of our proposed approach. The theoretical underpinning of our method is relegated to Appendix~\ref{sec:theory}.

\section{Related Work}
\label{sec:related-work}

We survey the immediate related literature and refer to \cite{choi2017online} and \cite{wang2017display} for more background on display advertising.

Most of the advertising is done through auctions, and we thus take the point of view of a bidder in this paper. Regardless of the auction mechanism, the bidder has to estimate the value he could generate from each display opportunity and then calibrate his bid accordingly. For example, in a second price setting, for a given display opportunity, the standard in the industry is to bid the estimated average value earned with this specific impression \citep{perlich2012bid, wang2017display}. This is referred to as \emph{value based bidding} in~\cite{xu2016lift}. We would like to emphasize that this behavior is \emph{myopically} designed in the sense that the future policy is disregarded or equivalently that the bid is computed as if this was the last opportunity to generate a conversion. Our approach to the label attribution problem adopts a similar philosophy and offers a principled myopic optimal approach. Without loss of generality, we assume that the bidder is compensated proportionally to the reward, i.e., the advertiser's attribution, that he received.
Note that the bidder might need to compute some shading factor to adapt to the competition and the auction mechanism \citep{balseiro2019learning,balseiro2021budget} or solve a broader optimization problem to reach certain requirements given a budget constraint~\cite{hojjat2017unified}. However, these computations are beyond the scope of this paper.

Realizing the need for more principled solutions to the advertiser attribution problem, recent works have started proposing algorithmic approaches to go beyond rule-based heuristics such as last touch. Using different generative models, \cite{Danaher2018desilusion} and \cite{anderl2016mapping} both propose a counterfactual approach. The notion of Shapley value~\citep{Berman2015,Besbes2019a,Dalessandro} has also been used to split the credit fairly among different channels. 
 \cite{Johnson2017a,Lewis2011a,Lewis2015} all study models related to causality and advertiser attribution. Their approach is based on econometric parametric models. 

Our research is inspired by the multi-touch attribution (MTA) literature~\citep{Shao2011, Ji2016, Li2018}. This stream of work aims at designing machine learning algorithms that assign fractional credits to different touchpoints from the advertiser's standpoint. They also aim at solving the advertiser attribution problem but focus on the implementability of the solution rather than its theoretical justification. Despite using a different formalism, our work relates to the survival analysis approach used in~\cite{Laub}. Moreover, the fixed point-based algorithm of Section~\ref{sec:additivity} shares some characteristics with the EM algorithm in \cite{Zhang2015c}. Similarly, \cite{Ji2017} relies on a Weibull model and is very close to our work. However, their choice of parametrization hides the fixed point property identified in~\cite{Zhang2015c}.

On top of the advertiser attribution problem, some researchers have also considered the bidder's problem as we do and tried to improve the buying process using machine learning pipelines.  \cite{Chapelle2014b} presents a bidding architecture, which relies on an implicit last touch label attribution.  \cite{Chapelle2014a} provides a technique to deal with the delay between displays and sales. \cite{diemert2017attribution} proposes to model the advertiser attribution  to increase bidding performances. Their experimental results motivate our approach.  More recently~\cite{bompaire2020causal} provides a framework based on reinforcement learning and causal inference.

Finally, we would like to highlight that the question of attribution is not specific to online advertising. For instance, recent work in interpretable machine learning tries to attribute the prediction of a machine learning model to individual features of the input \citep{dhamdhere2018important}. In a very different context, \cite{flores2020teamwork} develops a method to disentangle individual performance from team productivity. In each of these examples, the goal is to attribute some aggregate output to individual components of the input. Although \cite{flores2020teamwork} develops an iterative algorithm, which is similar in spirit to our approach, their problem of splitting credit among team members is static in nature, while time is an important aspect of our problem. Another important difference is that we propose a machine learning implementation of our algorithm, which is able to leverage the presence of context features.

\section{The Label Attribution Problem} \label{sec:model}

In this section, we formalize the label attribution problem. We begin by introducing the model and some notations.

\subsection{Model and notations}

In practice, a bidder is given a dataset of $T$ observations $\{\boldsymbol{x}^1, \dots, \boldsymbol{x}^T\}$, where for each $t \leq T$, $\boldsymbol{x}^t$ represents a display opportunity. Due to our focus on display advertising, we refer to an element $\boldsymbol{x}$ as a display or display opportunity in the paper. However, these could in principle represent any types of marketing interactions. We let $\mathcal{X} \subseteq \mathbb{R}^d$ be a set of potential display opportunities. Each $\boldsymbol{x} \in \mathcal{X}$ is a vector that encodes all the features associated with a display that the bidder can use to evaluate the display opportunity. For instance, this could include the type of products or creatives as well as contextual information such as whether or not the user has already seen an ad. In particular, we can also capture customer heterogeneity by encoding the user's features within $\boldsymbol{x}$. In general, any feature that is useful for the bidder can be encoded in the display set~$\mathcal{X}$.

\paragraph{Reward.} Additionally, each display $\boldsymbol{x}$ is associated with a reward. Most often, this reward represents whether the display lead to a conversion or not. Importantly, we assume that this reward is given for a subset of displays. For ease of exposition, we assume here that a group of display corresponds to the interaction with a specific user. However, nothing prevents the bidder to define a different grouping. More specifically, we assume that each display is associated with a user $u^t$. We denote by $\mathcal{U}$ the set of users. Each user $u \in \mathcal{U}$ is associated with a reward $r_u$. By doing this, we recognize that the reward $r_u$ is  generated by the set of displays $\mathcal{X}_u = \{ \boldsymbol{x}^t : \forall t, u^t=u \}$. From the bidder perspective, the value of the reward comes from a black-box decided by the advertiser. In the case where the bidder and advertiser are the same, these rewards could represent sales or any objective the advertiser is optimizing for. However, in general, the bidder and advertiser are different actors.

\paragraph{Valuation.} The bidder needs to estimate the value for each new display opportunity. We capture this using a valuation $V : \mathcal{X} \to \RR^+$, which  gives the value of a display opportunity. The bidder's challenge is to appropriately learn a valuation $\Valuation$ from the reward samples, given at the user level.  

\subsection{Label attribution} \label{subsec:label_attribution}

Since the rewards are given for each user, the reward data cannot be used directly to learn $V$. In order to train a ML algorithm, a training set has to be constructed from the reward data to learn $V$. This key step, which we call the label attribution, is often overlooked. It is nevertheless crucial, as different label attribution schemes lead to different valuation functions. One of our key contribution is to formalize this step.

\begin{definition}[label attribution]
\label{def:labell-attribution}
A \textbf{label attribution} $\Mapping$ is a  mapping from  $\mathcal{X} \times \mathcal{U}$ to  $\RR^+$ that satisfies the following properties.
\begin{enumerate}
    \item  $\forall u \in \mathcal{U}$ and $\forall \boldsymbol{x} \notin \mathcal{X}_u$, $\Mapping(\boldsymbol{x}|u) = 0$.
    \item $\forall u \in \mathcal{U}$, $r_u = \sum_{\boldsymbol{x} \in \mathcal{X}_u}  \Mapping(\boldsymbol{x}|u)$.
\end{enumerate}
\end{definition}

\medskip

For a label attribution $\mu$ and display $\boldsymbol{x} \in \mathcal{X}$, $\mu(\boldsymbol{x}|u)$ represents the label of the  display $\boldsymbol{x}$ for user $u$, i.e., the part of the reward $r_u$ attributed to display $\boldsymbol{x}$. The first property states that we can only split the reward of a user among its displays. The second property states that the attribution has to be balanced and does not attribute more than the reward $r_u$.  Just like the advertiser's attribution, the label attribution is splitting credit. However, the split is done at the more granular level of the displays. Indeed, the bidder needs to predict the value of a display opportunity in order to calibrate his bid in real-time auctions.  With this definition, a typical flow can be described as follows:

\medskip
\begin{framed}
\begin{enumerate}
\item \textbf{Label attribution step.} The label attribution $\mu$ is used to create a  training set $$S = \left \{ (\boldsymbol{x}^1, y^1) ,\dots, (\boldsymbol{x}^T,y^T) \right\},$$ where for each $t$, $y^t = \mu(\boldsymbol{x}^t| u^t)$.

\item \textbf{Prediction step.} Some ML algorithm, typically a logistic regression, is trained to predict the display labels $y$ from the corresponding vectors of features $\boldsymbol{x}$ and the resulting mapping constitutes the estimated valuation $\hat V$.
\end{enumerate}
\end{framed}
\medskip

As a concrete example, we can revisit the typical last touch approach with our notation. In particular, the last touch label attribution can be written for all $(\boldsymbol{x},u) \in \mathcal{X} \times \mathcal{U}$ as 
\begin{align*}
    \Mapping^{\sf LT}(\boldsymbol{x}|u) =\left \{ \begin{array}{cl}
    r_u & \text{, if } \boldsymbol{x} = \boldsymbol{x}^{\sf LT}_u, \\
    0 & \text{, otherwise, }
    \end{array} \right. 
\end{align*}
where $\boldsymbol{x}^{\sf LT}_u \in \mathcal{X}_u$ is the last display shown to user $u$. 

We would like to end this section by emphasing a key difference between the valuation and the label attribution. On the one hand, the label attribution is backward looking, i.e., it is a function of the entire set of displays a user has been exposed to. On the other hand, the valuation is forward-looking and only has access to the current display opportunity. Indeed, the valuation is to be used for bidding purposes.

\section{The Additive Valuation} \label{sec:additivity}

In this section, we propose a new approach to solve the label attribution problem, which is more principled and can efficiently be implemented in practice.

\subsection{Additivity and fixed point equation} \label{sec:fixed_point}

We begin by arguing for an intuitive property that we wish our valuation to satisfy. In particular, a very natural way to evaluate the next display opportunity is through its marginal contribution. This method has recently received attention in the industry~\citep{perlich2012bid, wang2017display}. It is also sometimes referred to as lift-based valuation \citep{xu2016lift}. We formalize this property using a notion of additivity for our valuation function.
\begin{definition}[Additivity] \label{def:additivity}
Let $\Valuation$ be a valuation. We say that $\Valuation$ satisfies the additivity property if for all $u \in \mathcal{U}$ and $\boldsymbol{x} \in \mathcal{X}_u$, $$\Valuation (\boldsymbol{x}) = r_{u'} - r_u,$$
where $\mathcal{X}_{u'} = \mathcal{X}_u \cup \{ \boldsymbol{x} \}$.
\end{definition}
In other words, the value associated with $\boldsymbol{x}$ is the incremental reward. Indeed, in the definition, users $u'$ and $u$ have been exposed to the same set of display ads expect for $\boldsymbol{x}$, which only user $u'$ has seen. The additivity property hinges on a temporal marginality, i.e., we are quantifying the marginal contribution with respect to the previous actions, and is in line with several works on the additivity effect of ads~\citep{Shao2011,Ji2017,Zhang2015c,Dalessandro}. This is different than a counterfactual marginality, used for instance in \cite{Danaher2018desilusion,anderl2016mapping,bompaire2020causal}, which quantifies the marginal contribution with respect to an alternative action. The counterfactual marginality depends on the policy used in the future, while the temporal marginality does not.

To learn such an additive valuation, we leverage a fixed point equation between the reward, the valuation and the label attribution. In particular, in an ideal setting where $V$ satisfies Definition~\ref{def:additivity} and is known, the label attribution would split the reward proportionally to $V$, i.e., for all $\boldsymbol{x} \in \mathcal{X}$ and $u \in \mathcal{U}$,
\begin{align} \label{eq:mtc_proof}
    \Mapping(\boldsymbol{x} | u) =  \frac{\Valuation\big(\boldsymbol{x}\big)}{\sum_{\boldsymbol{x}' \in \mathcal{X}_u} \Valuation\big( \boldsymbol{x'}\big)} \cdot r_u.
\end{align}
It is immediate to see that this is a valid label attribution. In Appendix~\ref{sec:theory}, we prove, using a slightly more abstract setting, that a valuation satisfies Definition~\ref{def:additivity} if and only if there exists a label attribution that satisfies Equation~\eqref{eq:mtc_proof} thereby formalizing the connection between proportional label attribution and additive valuation (see in particular Proposition~\ref{prop:fixed_point}). The theoretical justifications are relegated to the Appendix and we focus here on how to practically learn an additive valuation. In particular, we next show how to leverage the fixed point equation~\eqref{eq:mtc_proof} to design a learning algorithm. Before describing our algorithm, we introduce a measure of convergence, which is motivated by Definition~\ref{def:additivity}.

\subsection{A measure of convergence} \label{sec:likelihood}

As explained in the previous section, we want to develop an algorithm that learns an additive valuation, i.e., that is able to capture the lift in the reward provided by showing an ad. We use this intuition to propose a measure of convergence for our algorithm. More precisely, for every valuation $V$, we define $\mathcal{L}^{\sf add}$ as 
\begin{align*}
\mathcal{L}^{\sf add}(V) = \dfrac{1}{T} \cdot \sum \limits_{u \in \mathcal{U}}  \left[ r_u \cdot \ln \left( \sum_{\boldsymbol{x} : u^t = u} \Valuation( \boldsymbol{x}) \right)  - \sum_{\boldsymbol{x} : u^t = u} \Valuation (\boldsymbol{x})\right].
\end{align*}
The function $\mathcal{L}^{\sf add}$ can be thought of as a criterion to measure how well some valuation $V$ satisfies the additivity property. We use this loss to measure the progress of our algorithm, which we describe in the next section. Note that this loss resembles a likelihood function. However, $\mathcal{L}^{\sf add}$ is not per se a likelihood function since we do not have an underlying model whose parameters we are trying to find. In Appendix~\ref{app:additivity}, we show that under some assumptions, a valuation that satisfies Definition~\ref{def:additivity} minimizes $\mathcal{L}^{\sf add}(V)$ over all possible valuations.

\subsection{Fixed point label attribution (\textsf{FiPLA}) algorithm}

In Section~\ref{sec:fixed_point}, we introduce the fixed point equation~\eqref{eq:mtc_proof} relating the valuation, the label attribution and the reward. However, in practice, we do not know $V$. For that reason, we propose a  practical procedure that simultaneously learns $V$ and $\mu$. The proposed algorithm is iterative and maintains, in each iteration $k$, a label attribution $\Mapping^{(k)}$ and a valuation $\hat \Valuation^{(k)}$. Each iteration $k$ consists of three steps. 
\begin{itemize}
    \item The first two steps are a Label Attribution step and a Prediction step. These are similar to the one described in Section~\ref{subsec:label_attribution}.
    \item The third step is a fixed point step that exploits Equation~\eqref{eq:mtc_proof} to update the label attribution $\mu^{(k+1)}$ using the valuation computed in the previous step $\hat V^{(k+1)}$. 
\end{itemize} 
\medskip
Algorithm~\ref{alg:two} summarizes the procedure, which we refer to in the rest of the paper as the Fixed Point Label Attribution (\textsf{FiPLA}) Algorithm.  
\begin{algorithm}[t]
	\SetAlgoNoLine
	\KwIn{A dataset of displays $\{\boldsymbol{x}^1, \dots, \boldsymbol{x}^T\}$, a set of users $\mathcal{U}$ and associated rewards $r_1, \dots, r_{|\mathcal{U}|}$ and an initial attribution $\Mapping^{0}$}
	\KwOut{A valuation $\hat{V}^{\sf FP}$}
 \While{$|\mathcal{L}^{\sf add}(\hat{V}^{k+1}) - \mathcal{L}^{\sf add}(\hat{V}^{k})| \geq \epsilon$}{
    \vspace{0.2cm}
    1. \textit{Attribution step}.  Generate training dataset by  attributing rewards at the display level with the label attribution $\mu^k$
    \begin{align*}
       \mathcal{S}^{(k)} \;\; = \;\; \big\{
       (\boldsymbol{x}^1, y^1) ,\dots, (\boldsymbol{x}^T,y^T)
       \big\},
       \end{align*}
where for each $t$, $y^t = \mu^k(\boldsymbol{x}^t| u^t)$. 
      
    2. \textit{Prediction step}. Train any \textit{ML algorithm} on the training set $\mathcal{S}^{(k+1)}$. Update $\hat \Valuation^{(k+1)}$ to the resulting mapping. 
    
    3. \textit{Fixed point update}. Update the label attribution $\Mapping^{(k+1)}$ with the fixed point characterization 
       $$\mu^{(k + 1)} \;\; = \;\; (\boldsymbol{x}^t | u^t) \mapsto  \frac{\hat \Valuation^{(k+1)}\big(\boldsymbol{x}^t\big)}{\sum_{\boldsymbol{x}' \in \mathcal{X}_{u^t}} \hat \Valuation^{(k+1)}  \big(\boldsymbol{x}'\big)} \cdot r_{u^t}.$$
 }
	\caption{Fixed Point Label Attribution (\textsf{FiPLA}) Algorithm}
	\label{alg:two}
\end{algorithm}
Compared to existing approaches, the main novelties of our approach are (1) explicitly doing the label attribution (Step~1) instead of implicitly using a last touch label attribution, and (2) adding a fixed point update (Step 3). As also illustrated in Figure~\ref{fig:paradigm-change} in the Introduction, our approach suggests a change in paradigm whereby instead of estimating the valuation from a fixed internal procedure, such as last touch attribution, the optimal pair of valuation and label attribution are rather iteratively and jointly updated until convergence. Moreover, this feedback loop is a simple change that can easily be implemented in an existing ML pipeline.

Finally, we provide theoretical justifications to our algorithm and prove in Proposition~\ref{prop:convergence} of Appendix~\ref{sec:theory} that the procedure converges, under some assumption. This is done by relating our iterative procedure to a majorize-minorize algorithm.  

\section{Synthetic Experiments} \label{sec:numerics_synthetic}

In this section, we present some numerical experiments using synthetic data to build some intuition. The goal of this controlled experiment setup is to illustrate the benefits of using the \textsf{FiPLA} Algorithm over the last touch approach. 

\subsection{Experimental setups}

We present three different settings, each with a different generative process. 
\begin{enumerate}
\item \textbf{Constant valuation.} In this setup, we construct user timelines as follows. At each time step, the user is exposed to an ad and converts with a constant probability $\alpha$. Additionally, independent of everything else, at each time step, there is also a constant probability of $\beta$ that the user leaves the system. For each user timeline, we let the reward be equal to the number of conversions. In this case, the only feature that we use is the number $x$ of impressions, which in this case is also the length of the user timeline. Note that, in expectation, the reward grows proportionally to the number of impressions and we expect $V^{\sf FP}(x)$ to be constant equal to $\alpha$.
\item \textbf{Decreasing marginal return.} In this second experiment, the probability of getting a reward is still increasing with the number of ads seen, but we  now assume that it exhibits a decreasing marginal return effect. In particular, we use the same setup as previously except in the way we compute the reward. More precisely, for this setting, we associate a reward of 1 to a user's timeline if the user converted at least once. In this case, the reward grows exponentially to one as the number of impressions increases and we expect $V^{\sf FP}(x)$ to decay like $(1-\alpha)^x$. This aims at capturing a decreasing marginal return effect of advertising~\citep{Chapelle2014b,diemert2017attribution,Zhang2015c}. 
\item \textbf{Two display types.} In this third setup, we revisit the motivating example from Section~\ref{sec:motivating_example} and illustrate a multi-dimensional feature space. We assume that in each time step there is an impression for a type $A$ display with probability 0.5 and an impression for a type $B$ display otherwise. Building on Setup 1, we assume that the user converts with a constant probability $\alpha_A$ after a display of type $A$ and $\alpha_B$ after a display of type $B$. Independent of everything else, there is also a constant probability $\beta$ that the user leaves the system in each time step.  Let $x$, resp. $y$, be the number of displays of type A, resp. type $B$, a user has been exposed to.  Additionally, we let $z \in \{ A , B \}$ denote the type of the current display opportunity. We apply the \textsf{FiPLA} Algorithm with $\mathcal{X}  = \{ (x,y,z) : x \in \mathbb{N}, y \in \mathbb{N}, z \in \{A,B\}\}$. In this case, generalizing the first setup,  we expect $V^{\sf FP}(x,y,A)$ (resp. $V^{\sf FP}(x,y,B)$) to be constant equal to $\alpha_A$ (resp. $\alpha_B$) for all values of $x$ and $y$.
\end{enumerate}

\subsection{Results}

In each experiment, we generate timelines for 3M users and compare the valuation function $\hat{V}^{\sf LT}$ learned using a last touch approach with the valuation function $\hat{V}^{\sf FP}$  learned with the \textsf{FiPLA} Algorithm. Figure~\ref{fig:iter} shows how the loss progresses with the number of iterations for Setup~1. This is to illustrate that the number of iterations needed to run the \textsf{FiPLA} Algorithm in practice remains very reasonable. Even though we do not show these plots for all our experiments, the number of iterations remains similar throughout all the experiments of this section. 
\newcommand{\likelihoodA}{
( 0 , -0.6578513488043939 )
( 1 , -0.6119398142997413 )
( 2 , -0.6059358007512732 )
( 3 , -0.6046216370117375 )
( 4 , -0.6042575304032525 )
( 5 , -0.6041374958927387 )
( 6 , -0.6040911800050109 )
( 7 , -0.6040706517526493 )
( 8 , -0.604060491084136 )
( 9 , -0.6040550294489506 )}

\begin{figure}[h!]
\centering
 \begin{tikzpicture}[scale=0.8]
        \begin{axis}[xlabel={Iteration number $k$}, ylabel={}, axis y line=left, axis x line=bottom, xmin=0, xmax=7, legend pos=south east,
        grid=both,
    grid style={line width=.1pt, draw=gray!10, style=dashed},
    major grid style={line width=.2pt,draw=gray!50},ymax=-0.60, ymin = -0.66,
    ytick={-0.66,-0.65,-0.64,-0.63,-0.62,-0.61,-0.60},
    yticklabels={-0.66,-0.65,-0.64,-0.63,-0.62,-0.61,-0.60}]
        \addplot[blue,very thick] coordinates {\likelihoodA};
        \addlegendentry{$\mathcal{L}^{\sf add}(\hat{V}^k)$}
        \end{axis}
\end{tikzpicture}
  \label{fig:experiment_1_loss}
\caption{Convergence of the \textsf{FiPLA} Algorithm for Setup~1.}
\label{fig:iter}
\end{figure}

\paragraph{Setup 1.} Figure~\ref{fig:experiment_1} shows the learned valuations  $\hat{V}^{\sf LT}(x)$ and $\hat{V}^{\sf FP}(x)$ for Setup~1. We observe that the \textsf{FiPLA} Algorithm correctly learns the underlying true valuation since $\hat{V}^{\sf FP}(x)$ is constant equal to $\alpha = 0.1$. On the other hand, we see that $\hat{V}^{\sf LT}(x)$ predicts a higher valuation when $x$ is higher. This is precisely what we eluded to in the motivating example where the last touch approach tends to overestimate the value of impressions that appear later in the user journey and underestimate those that appear earlier. Indeed, if the label attribution is driven by the last touch philosophy, then an impression is evaluated through its probability of being the last impression. This is why we observe that $\hat{V}^{\sf LT}(x)$ increases with $x$ whereas the true underlying value of each impression remains constant.    

\newcommand{\likelihoodB}{
( 0 , -0.6578513488043939 )
( 1 , -0.6119398142997413 )
( 2 , -0.6059358007512732 )
( 3 , -0.6046216370117375 )
( 4 , -0.6042575304032525 )
( 5 , -0.6041374958927387 )
( 6 , -0.6040911800050109 )
( 7 , -0.6040706517526493 )
( 8 , -0.604060491084136 )
( 9 , -0.6040550294489506 )}
\newcommand{\lasttouchconstant}{
( 1 , 0.03003366666836805 )
( 2 , 0.059912759086877855 )
( 3 , 0.09000240138604462 )
( 4 , 0.11905611176545862 )
( 5 , 0.1499277471704117 )
( 6 , 0.18003800054428387 )
( 7 , 0.21021990586363987 )
( 8 , 0.23959549286559512 )
( 9 , 0.2679861638218139 )
( 10 , 0.2988944241232236 )
( 11 , 0.3336405638866117 )
( 12 , 0.3600426266115943 )
( 13 , 0.39004830917849115 )
( 14 , 0.41926257753267543 )
( 15 , 0.45227295138545676 )
( 16 , 0.48275132275124516 )
( 17 , 0.5192695722356567 )
( 18 , 0.5420492639556239 )
( 19 , 0.5752487562188994 )
( 20 , 0.5910703725605911 )
( 21 , 0.6475548060708018 )
( 22 , 0.6442831215971109 )
( 23 , 0.7227979274611295 )
( 24 , 0.7026022304832669 )
( 25 , 0.7731958762886524 )
( 26 , 0.8450000000000059 )
( 27 , 0.973684210526322 )
( 28 , 0.9132947976878627 )
}
\newcommand{\MTCconstant}{
( 1 , 0.0980326032926305 )
( 2 , 0.10187864850961167 )
( 3 , 0.10126164369044538 )
( 4 , 0.09903603092711293 )
( 5 , 0.10019444727261638 )
( 6 , 0.1000659253526741 )
( 7 , 0.09956622647315462 )
( 8 , 0.09835412966276162 )
( 9 , 0.09802823267818858 )
( 10 , 0.09997522552645274 )
( 11 , 0.1022662502683172 )
( 12 , 0.09957969446016368 )
( 13 , 0.09947713271924032 )
( 14 , 0.09828803710450049 )
( 15 , 0.10069263797662127 )
( 16 , 0.10078589183263398 )
( 17 , 0.10151525411765931 )
( 18 , 0.10409588686695753 )
( 19 , 0.10220564235009987 )
( 20 , 0.09713551281309388 )
( 21 , 0.10619426866565937 )
( 22 , 0.09745154520851214 )
( 23 , 0.11541048512721691 )
( 24 , 0.1000253239107007 )
( 25 , 0.09186748450500967 )
( 26 , 0.09769562796034131 )
( 27 , 0.12063016460320158 )
( 28 , 0.09940109885371128 )
( 29 , 0.09956482357255443 )
( 30 , 0.05283744372087328 )
( 31 , 0.0585323295729889 )
( 32 , 0.04685468352180738 )
( 33 , 0.1802709427892174 )
}

\newcommand{\lasttouchb}{
( 1 , 0.03003366666836805 )
( 2 , 0.05689709449158633 )
( 3 , 0.0813118246918127 )
( 4 , 0.102534471515983 )
( 5 , 0.12292004158806583 )
( 6 , 0.14083430452497642 )
( 7 , 0.15656412336880499 )
( 8 , 0.17078064202356774 )
( 9 , 0.18298173137713217 )
( 10 , 0.1945846642726582 )
( 11 , 0.20868044477537676 )
( 12 , 0.21548428566577232 )
( 13 , 0.22355072463767223 )
( 14 , 0.23328738800838678 )
( 15 , 0.2394734247115398 )
( 16 , 0.24458553791884027 )
( 17 , 0.2585754640839237 )
( 18 , 0.25331584317154926 )
( 19 , 0.26368159203980207 )
( 20 , 0.2637492607924448 )
( 21 , 0.2702360876897161 )
( 22 , 0.26678765880217353 )
( 23 , 0.27979274611398747 )
( 24 , 0.25774473358116856 )
( 25 , 0.2835051546391763 )
( 26 , 0.31499999999999845 )
( 27 , 0.3383458646616525 )
( 28 , 0.2947976878612711 )
( 29 , 0.3389830508474576 )
( 30 , 0.2564102564102559 )
( 31 , 0.3157894736842104 )
( 32 , 0.28947368421052627 )
( 33 , 0.3703703703703703 )
( 34 , 0.2941176470588235 )
( 35 , 0.4166666666666667 )
}
\newcommand{\MTCb}{
( 1 , 0.09795066329954477 )
( 2 , 0.09175686083442683 )
( 3 , 0.08223372329668287 )
( 4 , 0.07251987836246697 )
( 5 , 0.06586344145998552 )
( 6 , 0.05923021775477666 )
( 7 , 0.05279185113287734 )
( 8 , 0.04728549519300751 )
( 9 , 0.04259302769883315 )
( 10 , 0.03889942085113128 )
( 11 , 0.036336323546495304 )
( 12 , 0.03180917907666489 )
( 13 , 0.0287907276067532 )
( 14 , 0.02637728534901575 )
( 15 , 0.023727550605698148 )
( 16 , 0.021570476045434467 )
( 17 , 0.020709901064030943 )
( 18 , 0.01797587416821575 )
( 19 , 0.01695061658162333 )
( 20 , 0.014551760900188178 )
( 21 , 0.01329875215265713 )
( 22 , 0.01185555410683992 )
( 23 , 0.011819709954362463 )
( 24 , 0.009599501069029513 )
( 25 , 0.009564368819776308 )
( 26 , 0.010629851449775827 )
( 27 , 0.010615060509340243 )
( 28 , 0.00787993965313573 )
( 29 , 0.009399643599552775 )
( 30 , 0.005745323619163259 )
( 31 , 0.006906231994421822 )
( 32 , 0.0066146254861429235 )
} \newcommand{\likelihoodtwo}{
( 0 , -1.1280242464023986 )
( 1 , -1.0725425224457803 )
( 2 , -1.050763255203819 )
( 3 , -1.0388849921648613 )
( 4 , -1.0313660360571213 )
( 5 , -1.0261696943052445 )
( 6 , -1.0223618665465763 )
( 7 , -1.0194514034725881 )
( 8 , -1.0171550388315715 )
( 9 , -1.0152975649420342 )
( 10 , -1.013764678990165 )
( 11 , -1.0124786535711048 )
( 12 , -1.0113847462513232 )
( 13 , -1.0104432934345773 )
( 14 , -1.0096248541858774 )
( 15 , -1.0089071078777703 )
( 16 , -1.0082727824538396 )
( 17 , -1.0077083563202494 )
( 18 , -1.0072031225801301 )
( 19 , -1.006748350555224 )
( 20 , -1.0063370749969367 )
( 21 , -1.0059634576973904 )
( 22 , -1.0056227113626781 )
( 23 , -1.0053107655261344 )
( 24 , -1.0050243351194421 )
( 25 , -1.0047604823365328 )
( 26 , -1.0045167615716049 )
( 27 , -1.0042910513172445 )
( 28 , -1.004081552468342 )
( 29 , -1.003886681095719 )}
\newcommand{\lasttouchtwo}{
( 1.0 , 0.2687433189183753 )
( 2.0 , 0.259618680698409 )
( 3.0 , 0.25199179947620004 )
( 4.0 , 0.24041216329096068 )
( 5.0 , 0.23160259155419977 )
( 6.0 , 0.21941433990893908 )
( 7.0 , 0.20653375669849702 )
( 8.0 , 0.19175805957525982 )
( 9.0 , 0.17755065899115355 )
( 10.0 , 0.16254429225097614 )
( 11.0 , 0.14850535024487108 )
( 12.0 , 0.13494339949097225 )
( 13.0 , 0.12414444138780374 )
( 14.0 , 0.11857083844437659 )}
\newcommand{\iterationfivetwo}{
( 1.0 , 0.5055850056371249 )
( 2.0 , 0.3081553638274457 )
( 3.0 , 0.17828124145180282 )
( 4.0 , 0.09895817750998957 )
( 5.0 , 0.053718850139437946 )
( 6.0 , 0.02874206182154284 )
( 7.0 , 0.015271606069084832 )
( 8.0 , 0.008084238708637559 )
( 9.0 , 0.004272343916182222 )
( 10.0 , 0.002255726497969865 )
( 11.0 , 0.0011904953299327855 )
( 12.0 , 0.0006281609097958279 )
( 13.0 , 0.00033141884770007375 )
( 14.0 , 0.0001748528506092856 )
}

\newcommand{\MTCtwo}{( 1.0 , 0.7505471356570965 )
( 2.0 , 0.24322584621812474 )
( 3.0 , 0.035505024881085465 )
( 4.0 , 0.004198185151039107 )
( 5.0 , 0.00048263874429536923 )
( 6.0 , 5.530497862301834e-05 )
( 7.0 , 6.334977818853961e-06 )
( 8.0 , 7.256172917458003e-07 )
( 9.0 , 8.311282866996911e-08 )
( 10.0 , 9.51980965957655e-09 )
( 11.0 , 1.0904065202976107e-09 )
( 12.0 , 1.2489602350794803e-10 )
( 13.0 , 1.4305689103151359e-11 )
( 14.0 , 1.6385849200710843e-12 )
}

\newcommand{\MTCmarginal}{( 0 , 0.5081852924962589 )
( 1 , 0.20762022276141945 )
( 2 , 0.03115225201495364 )
( 3 , 0.0036808072912510948 )
( 4 , 0.00042151392498278805 )
( 5 , 4.809657927754215e-05 )
( 6 , 5.485789501581919e-06 )
( 7 , 6.256680460602343e-07 )
( 8 , 7.135863578736834e-08 )
( 9 , 8.138583881759685e-09 )
( 10 , 9.282204232575003e-10 )
( 11 , 1.0586524144547993e-10 )
( 12 , 1.2074124912304005e-11 )
( 13 , 1.3770760864716951e-12 )
}

\newcommand{\lasttouchmarginal}{( 0 , 0.00925897820306587 )
( 1 , 0.006724783719285532 )
( 2 , 0.010074360162975804 )
( 3 , 0.009203923980948853 )
( 4 , 0.011945440926257378 )
( 5 , 0.011915240961914486 )
( 6 , 0.014641865870501009 )
( 7 , 0.014007818809958605 )
( 8 , 0.012205065936664733 )
( 9 , 0.019400241118495365 )
( 10 , 0.014082418920848161 )
( 11 , 0.007968174788077542 )
( 12 , 0.020406409435418812 )
( 13 , -0.0020941726893548424 )} 
    \begin{figure}[h!]
\centering
\begin{subfigure}{.45\textwidth}
  \centering
  \begin{tikzpicture}[scale=0.8]
        \begin{axis}[xlabel={Number of impressions $x$}, ylabel={}, axis y line=left, axis x line=bottom, xmin=0, xmax=10,ymin = 0,ymax = 0.35,legend pos=north west,
        grid=both,
    grid style={line width=.1pt, draw=gray!10, style=dashed},
    major grid style={line width=.2pt,draw=gray!50},
    ytick={0,0.10,0.20,0.30},
    yticklabels={0,0.10,0.20,0.30}] \addplot[smooth,orange,very thick] coordinates {\lasttouchconstant};
        \addlegendentry{$\hat{V}^{\sf LT}(x)$}
        \addplot[smooth,blue,very thick] coordinates {\MTCconstant};
        \addlegendentry{$\hat{V}^{\sf FP}(x)$}
        \end{axis}
\end{tikzpicture}
  \caption{Setup 1: Constant valuation.}
  \label{fig:experiment_1}
\end{subfigure}
\begin{subfigure}{.45\textwidth}
  \centering
  \begin{tikzpicture}[scale=0.8]
        \begin{axis}[xlabel={Number of impressions $x$}, ylabel={}, axis y line=left, axis x line=bottom, xmin=0, xmax=10,ymin = 0,ymax =0.35,legend pos=north west,
        grid=both,
    grid style={line width=.1pt, draw=gray!10, style=dashed},
    major grid style={line width=.2pt,draw=gray!50},
    ytick={0,0.10,0.20,0.30},
    yticklabels={0,0.10,0.20,0.30}] \addplot[smooth,orange,very thick] coordinates {\lasttouchb};
        \addlegendentry{$\hat{V}^{\sf LT}(x)$}
        \addplot[smooth,blue,very thick] coordinates {\MTCb};
        \addlegendentry{$\hat{V}^{\sf FP}(x)$}
        \end{axis}
\end{tikzpicture}
  \caption{Setup 2: Decreasing marginal return.}
  \label{fig:experiment_2}
\end{subfigure}\caption{Learned valuations in synthetic experiments for Setups 1 and 2. Here we use $\alpha = 0.1$ and $\beta = 0.3$. Note that we only show the results for $x$ less than $20$ since for larger values of $x$, we have less than 100 data points in the training set.}
\label{fig:experiment_12}
\end{figure}

\paragraph{Setup 2.} Figure~\ref{fig:experiment_2} shows the the learned valuations  $\hat{V}^{\sf LT}(x)$ and $\hat{V}^{\sf FP}(x)$ for Setup~2. In this case, recall that the underlying model implies that the true valuation is decreasing convex. In other words, the more ads a user have been exposed to, the less valuable the next ad is. This is exactly what $\hat{V}^{\sf FP}(x)$ learns as illustrated by the convex decreasing shape. On the other hand, $\hat{V}^{\sf LT}(x)$ is increasing,  clearly failing to capture the decreasing marginal return effect from the generative process.

\paragraph{Setup 3.} In this experiment, the feature space has dimension three and we need to learn $V(x,y,z)$, where $x$ is the number of displays of type A (the useful displays), $y$ is the number of displays of type $B$ (the useless displays) and $z$ is the type of the current display opportunity. Figure~\ref{fig:experiment_3} shows the results of the learned valuations.
\newcommand{\VoneA}{
(0, 0.19708644121913108)
(1, 0.1979003270524106)
(2, 0.19334793663909822)
(3, 0.1946282561966955)
(4, 0.1862375603963872)
(5, 0.21095577115836642)
(6, 0.19144913118974577)
(7, 0.24327948969848834)
(8, 0.13229254495456405)
(9, 0.06835184211160784)
(10, 0.6943358292544143)
}
\newcommand{\VfiveA}{
(0, 0.20106288758264695)
(1, 0.19743756657539235)
(2, 0.18973001003439147)
(3, 0.1904971419294486)
(4, 0.19224682224192657)
(5, 0.19931406437276888)
(6, 0.209691227941511)
(7, 0.18796309523996796)
(8, 0.17296679444453328)
(9, 0.26149066652152775)
(10, 0.12793694395397318)
}
\newcommand{\VtenA}{
(0, 0.1999854355936921)
(1, 0.1958677059690827)
(2, 0.1909913623046576)
(3, 0.1901073981174819)
(4, 0.1912036902642828)
(5, 0.1772744066192838)
(6, 0.17802968971470762)
(7, 0.18207079698653175)
(8, 0.2007010619985343)
(9, 0.14900363674331704)
(10, 0.2643603665887141)
}
\newcommand{\VLToneA}{
(0, 0.059937892754110676)
(1, 0.12065924269514931)
(2, 0.18043846694562368)
(3, 0.24170505840221007)
(4, 0.2894503546100954)
(5, 0.36963484945545383)
(6, 0.42333333333332507)
(7, 0.4897750511247494)
(8, 0.6744186046511589)
(9, 0.39175257731958696)
(10, 0.871794871794872)
}
\newcommand{\VLTfiveA}{
(0, 0.07520948722169864)
(1, 0.13642855585260077)
(2, 0.19528193206240094)
(3, 0.2518952446589488)
(4, 0.3119615512437089)
(5, 0.38359564164664073)
(6, 0.4531672170519413)
(7, 0.5054054054053987)
(8, 0.5323943661971922)
(9, 0.7152941176470571)
(10, 0.7465753424657505)
}
\newcommand{\VLTtenA}{
(0, 0.08949925363038097)
(1, 0.1508333851605566)
(2, 0.210685543523154)
(3, 0.2693132499360534)
(4, 0.3381432798177822)
(5, 0.3861405868581178)
(6, 0.431537478705222)
(7, 0.5036359371334215)
(8, 0.5771704180064158)
(9, 0.6052287581699274)
(10, 0.8827361563517937)
}

\newcommand{\VoneB}{
(0, 0.0504021827415339)
(1, 0.05740785708253096)
(2, 0.062470104750746464)
(3, 0.0661639764843896)
(4, 0.06641893180199197)
(5, 0.08628984801351508)
(6, 0.07451981871483494)
(7, 0.08465975595277547)
(8, 0.09217837259246069)
(9, 0.02335013889955477)
(10, 0.10255935537739311)

}
\newcommand{\VfiveB}{
(0, 0.0512922894894581)
(1, 0.05379932149813106)
(2, 0.05696520583283923)
(3, 0.0637753464397933)
(4, 0.07148413207638954)
(5, 0.0750415592973719)
(6, 0.06268146988732606)
(7, 0.08057972986740197)
(8, 0.04961448502060548)
(9, 0.07173647623470812)
(10, 0.13211608142482717)
}
\newcommand{\VtenB}{
(0, 0.04993765809477517)
(1, 0.05119419281440033)
(2, 0.05727792028477233)
(3, 0.06052885398305271)
(4, 0.06755030431060481)
(5, 0.06879046442910197)
(6, 0.07533663443909508)
(7, 0.07895701101476282)
(8, 0.08613174303262124)
(9, 0.11569104479395276)
(10, 0.0847763446544069)
}
\newcommand{\VLToneB}{
(0, 0.01513185876942708)
(1, 0.07552765925059748)
(2, 0.13532516052481425)
(3, 0.1926606943943606)
(4, 0.25435198789006536)
(5, 0.31973878758009494)
(6, 0.383771929824552)
(7, 0.42087542087541924)
(8, 0.4621848739495812)
(9, 0.47863247863247743)
(10, 0.6764705882352938)
}
\newcommand{\VLTfiveB}{
(0, 0.030514294188461744)
(1, 0.09114566302670535)
(2, 0.14974169052608624)
(3, 0.21162474051900326)
(4, 0.2737957416823144)
(5, 0.334175236033513)
(6, 0.3804363960219663)
(7, 0.45624532535528395)
(8, 0.5068627450980339)
(9, 0.504878048780485)
(10, 0.6375000000000017)
}
\newcommand{\VLTtenB}{
(0, 0.045316710673262005)
(1, 0.10466903441554735)
(2, 0.16566363285993957)
(3, 0.22503170174976456)
(4, 0.2893828590736311)
(5, 0.3440448991757008)
(6, 0.40276142566620293)
(7, 0.49660977320553956)
(8, 0.5013912075681836)
(9, 0.6117788461538546)
(10, 0.6676384839650191)
}     \begin{figure}[h!]
\centering
\begin{subfigure}{.45\textwidth}
  \centering
  \begin{tikzpicture}[scale=0.85]
        \begin{axis}[legend columns=2,xlabel={Number of ads $x$}, ylabel={}, axis y line=left, axis x line=bottom, xmin=0, xmax=5,ymax=0.4,ymin=0,
        grid=both,
    grid style={line width=.1pt, draw=gray!10, style=dashed},
    major grid style={line width=.2pt,draw=gray!50},
    ytick={0,0.05,0.10,0.15,0.20,0.25,0.30,0.35},
    yticklabels={0,0.05,0.10,0.15,0.20,0.25,0.30,0.35},
    legend style={at={(0.5,1.3)},anchor=north}] \addplot[smooth,orange,very thick] coordinates {\VLToneA};
        \addlegendentry{$\hat{V}^{\sf LT}(x,0,A)$}
        \addplot[smooth,blue,very thick] coordinates {\VoneA};
        \addlegendentry{$\hat{V}^{\sf FP}(x,0,A)$}
        \addplot[smooth,orange,dashed,very thick] coordinates {\VLTfiveA};
        \addlegendentry{$\hat{V}^{\sf LT}(x,1,A)$}
        \addplot[smooth,blue,dashed,very thick] coordinates {\VfiveA};
        \addlegendentry{$\hat{V}^{\sf FP}(x,1,A)$}
          \addplot[smooth,orange,dotted,very thick] coordinates {\VLTtenA};
        \addlegendentry{$\hat{V}^{\sf LT}(x,2,A)$}
        \addplot[smooth,blue,dotted,very thick] coordinates {\VtenA};
        \addlegendentry{$\hat{V}^{\sf FP}(x,2,A)$}
        
        \end{axis}
\end{tikzpicture}
  \caption{Learned valuation when $z = A$}
  \label{fig:experiment_3_A}
\end{subfigure}
\begin{subfigure}{.45\textwidth}
  \centering
  \begin{tikzpicture}[scale=0.85]
        \begin{axis}[legend columns=2,xlabel={Number of ads $x$}, ylabel={}, axis y line=left, axis x line=bottom, xmin=0, xmax=5,ymax=0.4,ymin=0,legend style={at={(0.5,1.3)},anchor=north},
        grid=both,
    grid style={line width=.1pt, draw=gray!10, style=dashed},
    major grid style={line width=.2pt,draw=gray!50},
    ytick={0,0.05,0.10,0.15,0.20,0.25,0.30,0.35},
    yticklabels={0,0.05,0.10,0.15,0.20,0.25,0.30,0.35}] \addplot[smooth,orange,very thick] coordinates {\VLToneB};
        \addlegendentry{$\hat{V}^{\sf LT}(x,0,B)$}
        \addplot[smooth,blue,very thick] coordinates {\VoneB};
        \addlegendentry{$\hat{V}^{\sf FP}(x,0,B)$}
        \addplot[smooth,orange,dashed,very thick] coordinates {\VLTfiveB};
        \addlegendentry{$\hat{V}^{\sf LT}(x,1,B)$}
        \addplot[smooth,blue,dashed,very thick] coordinates {\VfiveB};
        \addlegendentry{$\hat{V}^{\sf FP}(x,1,B)$}
        \addplot[smooth,orange,dotted,very thick] coordinates {\VLTtenB};
        \addlegendentry{$\hat{V}^{\sf LT}(x,2,B)$}
        \addplot[smooth,blue,dotted,very thick] coordinates {\VtenB};
        \addlegendentry{$\hat{V}^{\sf FP}(x,2,B)$}
        
        \end{axis}
\end{tikzpicture}
  \caption{Learned valuation when $z = B$}
  \label{fig:experiment_3_B}
\end{subfigure}\caption{Learned valuations in synthetic experiments for Setup 3. Here we use $\alpha_A = 0.20$, $\alpha_B = 0.05$ and $\beta = 0.1$.}
\label{fig:experiment_3}
\end{figure}
Looking at what the \textsf{FiPLA} Algorithm learns when $z=A$, i.e., the impression is a display of type A, we observe in Figure~\ref{fig:experiment_3_A} that $\hat{V}^{\sf FP}(x,y,A)$ correctly recovers the constant valuation from Setup~1. Additionally, we see that the valuation does not change as a function $y$, which is consistent with the generative model. When $z=B$, we observe in Figure~\ref{fig:experiment_3_B} that $\hat{V}^{\sf FP}(x,y,B)$ also recovers the right constant valuation. On the other hand, when the label attribution is done using last touch, we observe in Figure~\ref{fig:experiment_3_A} that the valuation $\hat{V}^{\sf LT}(x,y,A)$ increases with $x$ as in Figure~\ref{fig:experiment_1}. Furthermore, as illustrated in Figure~\ref{fig:experiment_3_B}, $\hat{V}^{\sf LT}$ is in fact almost the same whether $z=A$ or $z=B$ and therefore the last touch method does not differentiate the different types of ads.

\medskip 

After seeing that our approach correctly captures critical aspects of typical behaviors and overcomes last touch weaknesses, we take it next to a real dataset.

\section{``Criteo Attribution Modeling for Bidding'' dataset} \label{sec:numerics_criteo}

In this section, we apply our framework to a real dataset with a large feature space. The feature space size is very high and each input is unique with very high probability,  as it is often the case in real data sets. In this type of contexts, we often rely on ML techniques. 

\subsection{Data description}
``The  dataset represents a sample of 30 days of Criteo live traffic data. 
Each line corresponds to one impression (a banner) that was displayed to a user. 
For each banner we have detailed information about the context, if it was clicked, if it led to a conversion and if it led to a conversion that was attributed to Criteo or not. Data has been sub-sampled and anonymized so as not to disclose proprietary elements'' \citep{diemert2017attribution}. In total, the dataset contains approximately 16M displays shown to 6M users among which 180K have converted. Each display has about 20 features including 10 categorical features describing contextual information with user-specific information in particular.
The nature of these categorical features is not disclosed but they are meant to be ``used to learn the click/conversion models''. We thus consider them as correct descriptors of the user's features at the time of the impression. This reflects real settings in which feature engineering is made to transform non structured user's history data into features that are suitable for a machine learning training.
Some other notable features include whether the display was clicked or not, the time since the last click and the number of displays clicked by this user (see~\cite{diemert2017attribution} for a complete description of the Criteo Attribution Modeling for Bidding Dataset).

\subsection{Preprocessing and implementation details}

\paragraph{Prepropressing.} In order to be closer in spirit to the standard last touch benchmark and have stronger signals, we only keep  displays that were clicked. Morever, to mitigate the  side effects caused by the fixed window of observation, we remove the users for which some clicks are missing since such information can be inferred from the data. Additionally, to compare our results with existing metrics (see ``Performance'' paragraph in Section~\ref{sec:numerics_discussion} below), we need binary rewards. Consequently, we split each user that triggered multiple conversions into multiple users. The resulting dataset contains 16M displays shown to 8M users among which 196K have converted.

\paragraph{Hashing trick.} For the prediction step in the \textsf{FiPLA} Algorithm, we use a regularized logistic regression together with a  hashing trick described in \cite{Chapelle2014b}. For completeness, we give more details on this procedure.
First, we treat all our variables as categorical features. To do so, we discretize the numerical features and  group values into buckets. In our dataset, there are only two numerical features: ``time since last click'' and ``number of clicked displays''. In logistic regression, categorical features are usually handled with one hot encoding where the size of the encoding is equal to the number of categories. For instance, suppose that there are three types of display, then the feature column corresponding to the display type is transformed as follows.
\begin{align*}
        \text{Display A} & \to [1,0,0] \\
        \text{Display B} & \to [0,1,0] \\
        \text{Display C} & \to [0,0,1] 
    \end{align*}
When the number of possible values for each categorical value explodes, this method is not scalable. This is where the hashing trick is useful since it allows mapping the input to a lower dimensional space of fixed and pre-determined size denoted by $m$. More precisely, this is done by using a hashing function $f$ that transforms the one-hot encoding vector to a lower dimensional vector. For our experiments, we have chosen $m=2^{13}=8,192$. This is similar in spirit to an auto-encoder in ML that learns a lower dimensional representation of the data. In this case, note that the mapping is not learned and there could potentially be better ways of projecting the data to a lower dimensional space. However, this method is standard in the industry and efficient enough to run meaningful experiments. Finally, note among the many implementations of the hashing trick that exist, we use the one implemented in \texttt{scikit-learn}~\citep{scikit-learn}.

\subsection{Results and discussion.} \label{sec:numerics_discussion} We use a 80/20 split on the user identifiers, where we train our model using 80 \% of the dataset and test on the remaining 20 \%.

\paragraph{Convergence.}
We  check  the convergence of our algorithm using $\mathcal{L}^{\sf add}(\Valuation^{(k)})$ as a criterion. Figure~\ref{fig:criteo_data_likelihood_convergence_train} shows that the procedure converges in a few iterations similar to the previous set of experiments. Figure~\ref{fig:criteo_data_likelihood_convergence_test} also illustrates that our solution  generalizes to the testing dataset. Note that since we use a regularized logistic regression in the prediction step of the \textsf{FiPLA} Algorithm, this introduces a small bias as can be seen by the non-monotonicity of the iterates in Figure~\ref{fig:criteo_data_likelihood_convergence}. Nevertheless, we observe a significant improvement compared to the last touch method.

\newcommand{\likelihoodconvergencetrain}{
(1, -0.087687445277575)
(2, -0.0865013989241172)
(3, -0.0864044865822292)
(4, -0.086424821821283)
(5, -0.0864553266135838)
(6, -0.0864784217742402)
(7, -0.0864963661417186)
(8, -0.086509126494012)
(9, -0.0865177193273544)
(10, -0.0865237183822311)
}

\newcommand{\likelihoodconvergencetest}{

(1, -0.0882591694946513)
(2, -0.087108410706362)
(3, -0.087025497965977)
(4, -0.0870506045649369)
(5, -0.0870834675830304)
(6, -0.0871105629678845)
(7, -0.087129295428408)
(8, -0.0871416961172226)
(9, -0.0871514080449517)
(10, -0.087158974545429)
} \begin{figure}[t]
    \centering
    \begin{subfigure}{.45\textwidth}
  \centering
  \begin{tikzpicture}[scale=0.75]
        \begin{axis}[xlabel={Iteration Number}, ylabel={$\mathcal{L}^{\sf add}$}, axis y line=left, axis x line=bottom, xmin=0.5, xmax=10.5,legend pos=north east,
        y label style={at={(-0.1,0.5)}},
        grid=both,
    grid style={line width=.1pt, draw=gray!10, style=dashed},
    major grid style={line width=.2pt,draw=gray!50},
    ytick={-0.0880, -0.0878, -0.0876, -0.0874, -0.0872, -0.0870, -0.0868, -0.0866, -0.0864 },
    yticklabels={-0.0880, -0.0878, -0.0876, -0.0874, -0.0872, -0.0870, -0.0868, -0.0866, -0.0864 },
    scaled y ticks=false,
    ymin=-0.088, ymax=-0.0862, legend pos=south east]
    \addplot[only marks,draw=blue, thick, mark=x, mark size = 5pt] coordinates
            {( 1 , -0.087687445277575 )};
            \addlegendentry{$\mathcal{L}^{\sf add}(\hat{V}^{\sf LT})$}
            \addplot[only marks,draw=red, thick, mark=o, mark size = 3pt] coordinates
            {( 10 , -0.0865237183822311 )};
            \addlegendentry{$\mathcal{L}^{\sf add}(\hat{V}^{\sf FP})$}
        \addplot[smooth,red,thick] coordinates {\likelihoodconvergencetrain};
        \end{axis}
\end{tikzpicture}
  \caption{Training set}
  \label{fig:criteo_data_likelihood_convergence_train}
\end{subfigure}
~
\begin{subfigure}{.45\textwidth}
  \centering
  \begin{tikzpicture}[scale=0.75]
        \begin{axis}[xlabel={Iteration Number}, ylabel={$\mathcal{L}^{\sf add}$}, axis y line=left, axis x line=bottom, xmin=0.5, xmax=10.5,legend pos=south east,
        y label style={at={(-0.1,0.5)}},
        grid=both,
    grid style={line width=.1pt, draw=gray!10, style=dashed},
    major grid style={line width=.2pt,draw=gray!50},
    ytick={-0.0884, -0.0882, -0.0880, -0.0878, -0.0876, -0.0874, -0.0872, -0.0870},
    yticklabels={-0.0884, -0.0882, -0.0880, -0.0878, -0.0876, -0.0874, -0.0872, -0.0870},
    scaled y ticks=false,
    ymin=-0.0884, ymax=-0.0868] \addplot[only marks,draw=blue, thick, mark=x, mark size = 5pt] coordinates
            {( 1 , -0.0882591694946513 )};
            \addlegendentry{$\mathcal{L}^{\sf add}(\hat{V}^{\sf LT})$}
            \addplot[only marks,draw=red, thick, mark=o, mark size = 3pt] coordinates
            {( 10 , -0.087158974545429 )};
            \addlegendentry{$\mathcal{L}^{\sf add}(\hat{V}^{\sf FP})$}
        \addplot[smooth,red,thick] coordinates {\likelihoodconvergencetest};
        
        \end{axis}
\end{tikzpicture}
  \caption{Testing set}
  \label{fig:criteo_data_likelihood_convergence_test}
\end{subfigure}
    \caption{Convergence $\mathcal{L}^{\sf add}(\Valuation^{(k)})$ on the Criteo Attribution dataset.}
    \label{fig:criteo_data_likelihood_convergence}
\end{figure}

To put these results in perspective, we additionally tested the sensitivity of our results with respect to the size of the feature space. Indeed, with the hashing trick, we are embedding our feature space  into a target space whose size can be controlled. Figure~\ref{fig:criteo_data_likelihood_given_hash_space} illustrates the value of $\mathcal{L}^{\sf add}(\hat{\Valuation}^{\sf FP})$ under different feature space sizes, where recall that $\hat{V}^{\sf FP}$ is the output of the \textsf{FiPLA} Algorithm. As expected, $\mathcal{L}^{\sf add}(\hat{\Valuation}^{\sf FP})$ increases with respect to the feature space size. On the other hand, there is a priori no reason to believe that when increasing the size of the feature space, $\mathcal{L}^{\sf add}(\Valuation^{\sf LT})$ also increases. Indeed,  the last touch method, unlike our valuation, is not designed to maximize $\mathcal{L}^{\sf add}$. Somewhat reassuringly, we observe that $\mathcal{L}^{\sf add}(\Valuation^{\sf LT})$ also increases with the feature space size. Interestingly, we observe that the improvement, as measured by $\mathcal{L}^{\sf add}$, of going from $V^{\sf LT}$ to $\hat{V}^{\sf FP}$ are similar in magnitude to increasing the size of the feature space by a factor five, as illustrated by Figure~\ref{fig:criteo_data_likelihood_given_hash_space}.

\newcommand{\coretrain}{
(1024, -0.0887845148233382)
(2048, -0.0876952915997205)
(4096, -0.0869932771734129)
(8192, -0.0865237183822311)
(16384, -0.0862010786225254)
(32768, -0.0859868792182674)
(65536, -0.0858794195392055)
}

\newcommand{\lasttouchtrain}{
(1024, -0.0897239218705325)
(2048, -0.0887545601637645)
(4096, -0.0881243571025794)
(8192, -0.08768547836086)
(16384, -0.0873821221268108)
(32768, -0.0871851905995666)
(65536, -0.087081106449842)
}

\newcommand{\coretest}{
(1024, -0.088740390457676)
(2048, -0.0878106883917024)
(4096, -0.0873284882312004)
(8192, -0.087158974545429)
(16384, -0.0870469423090367)
(32768, -0.0870170507416737)
(65536, -0.08698318292964)
}

\newcommand{\lasttouchtest}{
(1024, -0.0896559751627884)
(2048, -0.0888142073164815)
(4096, -0.0883887110127546)
(8192, -0.0882605130663084)
(16384, -0.0881687156751783)
(32768, -0.0881490027027049)
(65536, -0.0881193329732249)
} \begin{figure}[t]
        \centering
      \begin{tikzpicture}\begin{axis}[xlabel={Size of feature space}, ylabel={$\mathcal{L}^{\sf add}$}, axis y line=left, axis x line=bottom, xmin=0, xmax=70000, ymin=-0.09, ymax = -0.085, legend style={at={(1,0.85)},anchor=north east},
        y label style={at={(-0.05,0.5)}},
        grid=both,
    grid style={line width=.1pt, draw=gray!10, style=dashed},
    major grid style={line width=.2pt,draw=gray!50},
    scaled x ticks=true,
    ytick={-0.090,-0.089, -0.088, -0.087, -0.086},
    yticklabels={-0.090,-0.089, -0.088, -0.087, -0.086},
    scaled y ticks=false]
     
        \addplot[smooth,red,thick,mark=o, mark size = 2pt] coordinates {\coretest};
        \addlegendentry{$\mathcal{L}^{\sf add}(\hat{V}^{\sf FP})$}
        \addplot[smooth,blue,thick,mark=x, mark size = 3pt] coordinates {\lasttouchtest};
        \addlegendentry{$\mathcal{L}^{\sf add}(\hat{V}^{\sf LT})$}
        \addplot[smooth,blue,dashed] coordinates {( 1024 , -0.0896559751627884 ) ( 65536 , -0.0896559751627884 )};
        \draw [very thick,decoration={brace,raise =1pt, amplitude=5pt},decorate] (axis cs:65536 , -0.0896559751627884) -- node[left = 2pt ] {$\Delta_{\sf features}$}  (axis cs:65536 , -0.0881193329732249) ;
    \draw [very thick,decoration={brace, raise =1pt, amplitude=5pt},decorate] (axis cs:32768 , -0.0881490027027049) -- node[left = 2pt ] {$\Delta_{\sf add}$}  (axis cs:32768 , -0.0870170507416737) ;
        
        \end{axis}
\end{tikzpicture}
\caption{Additivity loss $\mathcal{L}^{\sf A}$ on the test set as a function of the features space size. $\Delta_{\sf add}$ represents the improvement from moving from $V^{\sf LT}$ to $\hat{V}$ when $m=2^{10}$. $\Delta_{\sf features}$ represents the improvement of $\mathcal{L}^{\sf add}(\hat{V}^{\sf LT})$ when going from $m=2^{10}$ to $m=2^{16}$.}
    \label{fig:criteo_data_likelihood_given_hash_space}
\end{figure}

\paragraph{Qualitative insights.} In the previous paragraph, we show that our method is improving $\mathcal{L}^{\sf add}$. However, this is more of a sanity check as our method is designed to maximize $\mathcal{L}^{\sf add}$. We next try to get some insights into how the proposed valuation is capturing some key elements of the problem. We discuss two important effects that have been discussed in the literature and recognized in practice. 

The first one is the notion that the effect of a click decays over time. Consequently, it is less valuable to place an ad immediately after another one has been clicked. Several works have tackled this problem by making some assumptions on the underlying mechanism at play  and hard-coding this decaying effect \citep{Chapelle2014b,diemert2017attribution,Zhang2015c}. On the contrary, we do not hard-code this relationship into our algorithm. Nevertheless, our valuation is able to learn this effect from data. Indeed, we show in Figure~\ref{fig:sanity_checks_time} the relative difference between $\hat{V}^{\sf FP}$ and $V^{\sf LT}$ as a function of the feature ``hours since last click''. We observe that the last touch valuation tends to overestimate the value of an opportunity when a display was just clicked and our approach is able to correct for this effect. This is coherent with the live experiments done in \cite{diemert2017attribution}.

Another effect that has been observed in practice is the diminishing marginal effect of ads. As a user is being exposed to different displays, the incremental benefit of showing an extra ad decreases. This is similar in spirit to the motivating example in the introduction. This is what we explore in Figure~\ref{fig:sanity_checks_clicks} where we show the relative difference between $\hat{V}^{\sf FP}$ and $V^{\sf LT}$ as a function of the feature ``number of clicks before display''. We observe that, compared to the last touch benchmark, our  framework drastically reduces the valuation of a display with the number of preceding clicks. 
\newcommand{\timeelapsed}{
(12, -0.5571380880873714)
(24, -0.5505496843980392)
(36, -0.5216662433452585)
(48, -0.5462805933512985)
(60, -0.527178371187066)
(72, -0.5147998221448483)
(84, -0.5135123623551282)
(96, -0.5124079498944271)
(108, -0.4885994379088771)
(120, -0.4790433135627686)
(132, -0.4564019500787266)
(144, -0.4371678010486001)
(156, -0.4166863708635429)
(168, -0.3934961122421236)
(180, -0.3772764064208593)
(192, -0.3484951283099183)
(204, -0.3038443721636027)
(216, -0.2862418587027438)
(228, -0.2670345302556013)
(240, -0.2723727060992044)
(252, -0.2758836820698134)
(264, -0.2577080442890182)
(276, -0.2332765324195188)
(288, -0.2182608261820101)
(300, -0.1773008158526313)
(312, -0.1580126346515531)
(324, -0.1750034092216449)
(336, -0.1439017758866343)
(348, -0.1358368447160734)
(360, -0.114087645193272)
(372, -0.1136324978204344)
(384, -0.0656887571095893)
(396, -0.0608920521456547)
(408, -0.0355065343246736)
(420, -0.0392976355082409)
(432, 0.0309294651221103)
(444, 0.049800689534487)
(456, 0.0652624941171401)
(468, 0.0861798969666646)
(480, 0.0882390397267731)
(492, 0.0886470428281857)
(504, 0.1200984943435894)
(516, 0.1269893908522305)
(528, 0.1511125555497571)
(540, 0.1582561966120511)
(552, 0.1920850989861068)
(564, 0.2034691862555147)
(576, 0.2407297816831508)
(588, 0.231121560186597)
(600, 0.262110101795397)
(612, 0.2669287170045141)
(624, 0.3149814182293639)
(636, 0.3388619661718697)
(648, 0.3543131611163726)
(660, 0.3608004346627428)
(672, 0.383431383246283)
(684, 0.3871961587890954)
(696, 0.4021327232228625)
(708, 0.4129959076277738)
(720, 0.6259112769311834)
}

\newcommand{\numberclicks}{
(0, 0.3107698020052015)
(1, -0.6109818090139786)
(2, -0.6963984846474122)
(3, -0.795078608548681)
(4, -0.8565746552541459)
(5, -0.8927810590965951)
(6, -0.9078254674981286)
(7, -0.9240569662907796)
(8, -0.8818924145025678)
(9, -0.9599002070293896)
} \begin{figure}[t]
    \centering
    \begin{subfigure}[t]{0.5\textwidth}
        \centering
       
       \begin{tikzpicture}[scale=0.75]
        \begin{axis}[xlabel={Hours since last click}, ylabel={Average valuation in relative difference}, axis y line=left, axis x line=bottom, xmin=0, xmax=700,legend pos=north east,
        y label style={at={(-0.05,0.5)}},
        grid=both,
    grid style={line width=.1pt, draw=gray!10, style=dashed},
    major grid style={line width=.2pt,draw=gray!50},
    scaled x ticks=true,
    yticklabel={\pgfmathparse{\tick*1}\pgfmathprintnumber{\pgfmathresult}\%},
     legend pos=north west]
        \addplot[smooth,red,thick] coordinates {\timeelapsed};
        \addlegendentry{Average}
        \end{axis}
\end{tikzpicture}
        \caption{Time elapsed since the last user's click}
         \label{fig:sanity_checks_time}
    \end{subfigure}~ 
    \begin{subfigure}[t]{0.5\textwidth}
        \centering
       \begin{tikzpicture}[scale=0.75]
        \begin{axis}[xlabel={Number of clicks before display}, ylabel={Average valuation in relative difference}, axis y line=left, axis x line=bottom, xmin=0, xmax=7,legend pos=north east,
        y label style={at={(-0.05,0.5)}},
        grid=both,
    grid style={line width=.1pt, draw=gray!10, style=dashed},
    major grid style={line width=.2pt,draw=gray!50},
    scaled x ticks=true,
    yticklabel={\pgfmathparse{\tick*1}\pgfmathprintnumber{\pgfmathresult}\%},
     legend pos=north east]
        \addplot[smooth,red,thick] coordinates {\numberclicks};
        \addlegendentry{Average}
        \end{axis}
\end{tikzpicture}
    \caption{Number of clicks already made by the user}
     \label{fig:sanity_checks_clicks}
    \end{subfigure}
    \caption{Relative variation of the display valuation between $\hat{V}^{\sf FP}$ and $V^{\sf LT}$.}
    \label{fig:sanity_checks}
\end{figure}

\paragraph{Performance.}
\label{par:performance}

Finally, we also evaluate the performance of our approach with a procedure used  in \cite{Zhang2015c} for Multitouch Attribution models.
This procedure consists in mapping the conversion probabilities of each display in a given user's history to a single user's conversion probability and was first proposed in \cite{Dalessandro} as a generative model.  The proposed conversion probability of a history can be written with our notations as follows
\begin{align} \label{eq:daless}
\mathbb{P}(R_u = 1) = \bigg( 1 - \prod_{t : u^t =u} \big( 1 - \Valuation ( \boldsymbol{x}^t))  \bigg) \cdot \delta^{|\mathcal{X}_u|}
\end{align}
Here, $\mathbb{P}(\Reward_u = 1)$ denotes the probability that a user $u$ converts. $R_u$ is a binary reward, and for each display  $\boldsymbol{x}^t \in \mathcal{X}_u$, we identify conversion probability of $\boldsymbol{x}^t$ with its valuation $V(\boldsymbol{x}^t)$. Additionally, $|\mathcal{X}_u|$ represents the number of display interactions with user $u$. The first term in brackets represents the probability that at least one display leads to a conversion assuming zero interaction effects. The second term $\delta^{|\mathcal{X}_u|}$ accounts for the marginally decreasing effect of each ad. \cite{Zhang2015c} fixes $\delta$ to 0.95. We use the same value for consistency. This metric aims at measuring how well the conversion for each displays can recover the reward of a timeline. We can then compare two models by evaluating how well they rank the converted users according to Equation~\eqref{eq:daless}. A natural metric in this case is the mean average precision metric since it summarizes the Precision-Recall curve used in \cite{Zhang2015c} with a single value. For that metric, our method provides an out-of-sample improvement of 20.5\% over the last touch attribution approach. This shows that our approach strongly outperforms the last touch valuation on this independent performance metric as well.

\section{Impact on revenue} \label{sec:numerics_revenue}

Properly addressing the label attribution problem helps better quantifying the value of a display opportunity. Ultimately, these values are used in downstream tasks such as bidding in a real-time bidding auction. In this section, we numerically explore the impact of the label attribution on the bidder's profit. More precisely, we simulate a  bidding environment and compare two bidding strategies. These two strategies differ in their solution to the label attribution problem and therefore in the valuation function they use. In particular, one uses the last touch label attribution approach whereas the other one uses the \textsf{FiPLA} Algorithm.
\subsection{Experiment setup}

\paragraph{Display types.} We revisit the synthetic experiments with an additional bidding layer. In particular, we assume that there are two types of displays. Displays of type $A$ lead to a conversion with a constant probability $\alpha$ whereas displays of type $B$ never trigger a conversion. Similar to Setup~2 in Section~\ref{sec:numerics_synthetic}, we associate a reward to a history if the user converts at least once, thus creating an effect of decreasing marginal returns when showing additional ads (of type $A$). We assume that the display type in each time step is sampled from a Bernoulli distribution of fixed parameter $0.5$.  As in Section~\ref{sec:numerics_synthetic}, at each time step, there is a constant probability $\beta=1/4$ that the user leaves the system. 

\paragraph{Bidding environment.} For each display opportunity, we assume that a second price auction takes place. More precisely, for each time step, we sample the highest bid $b^{\sf high}$ from a Beta distribution whose parameters depend on the display type. We assume that there is stronger competition for the displays of type $A$. 
More precisely, for display of type $A$ the competition was generated with a Beta distribution of parameters $(5,5)$, and for displays of type $B$, the competition was generated with a Beta distribution of parameters $(1,9)$.
 The bidder computes a bid $b$ using a given bidding strategy. If the $b$ exceeds $b^{\sf high}$, then the bidder wins the auction and pays $b^{\sf high}$. On the other hand, if $b$ is below $b^{\sf high}$, then the bidder looses the auction. For each user's timeline, the bidder generates a reward if the user converts at least once and incurs a cost for each won auction. 

\paragraph{Bidding strategies.} We compare two strategies that differ on their label attribution solution and therefore on their estimated valuation function $\hat{V}$, where recall that the valuation function estimates the value of a display opportunity. This estimated valuation is either $\hat{V}^{\sf LT}$, when the label attribution step is last touch, or $\hat{V}^{\sf FP}$ when the \textsf{FiPLA} Algorithm is used. We naturally refer to the two strategies as the \emph{last touch} and \textsf{FiPLA} strategies.
Both \emph{last touch} and \textsf{FiPLA} Algorithms were trained on the data generated by a bidder that always bids  $0.5$ on 10,000 users. We then evaluate both strategies on 10,000 new users.
In terms of bidding, as often in practice to control budget spend \citep{balseiro2021budget}, we do not directly use the value given by $V$ as a bid but allow the strategy to shade or modulate the value by a constant multiplier. For each strategy, we compute the revenue and spend for different values of bidding multiplier allowing us to generate a Pareto curve in the revenue-spend space.

\subsection{Results}

\newcommand{\paretocore}{
(117.41764756043266, 309)
(195.86776191588984, 470)
(294.72893302710463, 642)
(408.69572773067546, 821)
(563.5999786364052, 1073)
(732.685711091235, 1309)
(932.466036673319, 1574)
(1153.793920706693, 1827)
(1386.0132139159666, 2072)
(1642.2195469837015, 2334)
(1909.7474837871969, 2608)
(2166.700923448114, 2828)
(2417.394340361866, 3046)
(2684.46717624664, 3237)
(2942.99338487178, 3441)
}

\newcommand{\paretolasttouch}{
(23.614979972950582, 63)
(40.68692532123388, 90)
(60.61736697201259, 130)
(88.45451194851361, 173)
(136.27029165965263, 243)
(201.54813701516136, 326)
(305.00233739021166, 441)
(437.302930131068, 585)
(598.3882642074055, 732)
(777.5711925145481, 879)
(1003.6685941134195, 1055)
(1259.3037932073014, 1271)
(1561.454173587691, 1486)
(1879.90686112872, 1692)
(2270.609934933642, 1932)
(2632.815606150129, 2152)
(3060.3840482376136, 2410)
(3486.291421963973, 2662)
} \newcommand{\multipliercore}{
(0.5, 191.58235243956733)
(0.55, 274.13223808411016)
(0.6000000000000001, 347.27106697289537)
(0.6500000000000001, 412.3042722693245)
(0.7000000000000002, 509.4000213635948)
(0.7500000000000002, 576.314288908765)
(0.8000000000000003, 641.533963326681)
(0.8500000000000003, 673.2060792933071)
(0.9000000000000004, 685.9867860840334)
(0.9500000000000004, 691.7804530162985)
(1.0000000000000004, 698.2525162128031)
(1.0500000000000005, 661.2990765518862)
(1.1000000000000003, 628.6056596381341)
(1.1500000000000006, 552.5328237533599)
(1.2000000000000006, 498.0066151282199)
}

\newcommand{\multiplierlasttouch}{
(1.0, 39.38502002704941)
(1.1, 49.31307467876612)
(1.2000000000000002, 69.38263302798741)
(1.3000000000000005, 84.54548805148639)
(1.4000000000000004, 106.72970834034736)
(1.5000000000000004, 124.45186298483864)
(1.6000000000000003, 135.99766260978834)
(1.7000000000000006, 147.69706986893203)
(1.8000000000000007, 133.61173579259446)
(1.9000000000000008, 101.4288074854519)
(2.000000000000001, 51.331405886580455)
(2.100000000000001, 11.696206792698604)
(2.200000000000001, -75.454173587691)
(2.300000000000001, -187.9068611287196)
(2.400000000000001, -338.60993493364185)
(2.5000000000000013, -480.8156061501295)
(2.6000000000000014, -650.3840482376136)
(2.700000000000001, -824.2914219639729)
} \begin{figure}[t]
    \centering
    \begin{subfigure}[t]{0.5\textwidth}
        \centering
       
       \begin{tikzpicture}[scale=0.75]
        \begin{axis}[xlabel={spend}, ylabel={revenue}, axis y line=left, axis x line=bottom,legend pos=south west,
        grid=both,
    grid style={line width=.1pt, draw=gray!10, style=dashed},
    major grid style={line width=.2pt,draw=gray!50},
    xtick={0, 1000, 2000, 3000},
    xticklabels={\$0, \$1K, \$2K, \$3K},
    ytick={0, 1000, 2000, 3000},
    yticklabels={\$0, \$1K, \$2K, \$3K},
    scaled x ticks=true,
     legend pos=north west]
        \addplot[smooth,black,dashed] coordinates {( 0 , 0 ) ( 3000, 3000 )};
        \addlegendentry{profit $\geq 0$}
        \addplot[smooth,red,thick,mark=o, mark size = 2pt] coordinates {\paretocore};
        \addlegendentry{\textsf{FiPLA}}
        \addplot[smooth,blue,thick,mark=x, mark size = 3pt] coordinates {\paretolasttouch};
        \addlegendentry{Last touch}
        \end{axis}
\end{tikzpicture}
       \caption{Pareto curve for different multiplier values}
         \label{fig:simulation_pareto}
    \end{subfigure}~ 
    \begin{subfigure}[t]{0.5\textwidth}
        \centering
       \begin{tikzpicture}[scale=0.75]
        \begin{axis}[xlabel={Bidding multiplier}, ylabel={Profit (revenue - spend)}, axis y line=left, axis x line=bottom,legend pos=north east,
        xmin=0, xmax=3,
        ymin=-1200, ymax=1200,
        grid=both,
    grid style={line width=.1pt, draw=gray!10, style=dashed},
    major grid style={line width=.2pt,draw=gray!50},
    scaled x ticks=true,
    ytick={-1000, 0, 1000},
    yticklabels={-\$1K, \$0, \$1K},
     legend pos=north east]
        \addplot[smooth,red,thick,mark=o, mark size = 2pt] coordinates {\multipliercore};
        \addlegendentry{\textsf{FiPLA}}
        \addplot[smooth,blue,thick,mark=x, mark size = 3pt] coordinates {\multiplierlasttouch};
        \addlegendentry{Last touch}
        \end{axis}
\end{tikzpicture}
    \caption{Profit as a function of the bidding multiplier.}
     \label{fig:simulation_multiplier}
    \end{subfigure}
    \caption{Revenue and spend of both the \textsf{FiPLA} and last touch strategies for different bidding multipliers.}
    \label{fig:revenue_simulation_plots}
\end{figure}

Figure~\ref{fig:revenue_simulation_plots} illustrates the results for different multiplier values. Looking at Figure~\ref{fig:simulation_pareto}, we see that for both strategies, a higher spend leads to a higher revenue. Indeed, the multiplier is often adjusted in practice to match the spend with a budget. We observe however that the \textsf{FiPLA} strategy clearly dominates the last touch strategy. Indeed, for all spend values, the revenue generated by the \textsf{FiPLA} strategy strictly exceeds that of the last touch strategy. Moreover, the Pareto curve for the \textsf{FiPLA} strategy is always above the 45 degree line, indicating that, for all the multiplier values, the profit, which is the difference between the revenue and the spend, is always positive. On the other hand, the last touch strategy incurs a negative profit for some multiplier values. 

Inspecting the profit as a function of the multiplier, we see on Figure~\ref{fig:simulation_multiplier} that for the \textsf{FiPLA} strategy, the higher profit is obtained when the multiplier is one, i.e., when the valuation $\hat{V}^{\sf FP}$, is used directly as a bid. Note that this is not the case for the last touch approach, which needs to be carefully calibrated. Even allowing for the last touch strategy to calibrate a multiplier, we see, as summarized in Table~\ref{tab:my_label} that the profit generated by the \textsf{FiPLA} strategy is more than 3.5 times higher than the profit generated by the last touch strategy. This illustrates the importance of the label attribution problem and the benefits of properly addressing it.

\begin{table}[h]
    \centering
    \begin{tabular}{c|c c}
         & No multiplier & Best multiplier \\
         \hline \hline
Last touch strategy          & \$39 & \$148 \\
\textsf{FiPLA} strategy & \$698 & \$698\\
\hline 
    \end{tabular}
    \caption{Profit obtained by both strategies with and without bidding multipliers. Note that the \textsf{FiPLA} strategy does not benefit from using a bidding multiplier. This can be interpreted as a \textit{calibration} property, that is not satisfied by the last touch strategy.}
    \label{tab:my_label}
\end{table}

\section{Conclusion} \label{sec:discussion} 
In this paper, we introduce the label attribution problem which consists of splitting the reward given to a bidder for each users into labels for each display opportunity. This is an important step needed to create a training dataset for most ML algorithm but it is often overlooked in practice. We formalize this problem and propose an approach to this problem which we call the robust label attribution. Our method is motivated by an intuitive additivity property and enjoys several theoretical structural properties. It is also practical and we show how it can be implemented  at the scale required for display advertising.

\bibliographystyle{informs2014}

\newpage

\renewcommand{\theHsection}{A\arabic{section}}

\begin{APPENDICES}

\ECHead{\begin{center}
Appendix 
\end{center}
}

\section{Theoretical grounding for additive valuation} \label{sec:theory}

In this section, we provide some theoretical justification for our approach. To be able to prove our results, we first present a slightly more abstract setting.

\subsection{Model : from users to history}

The goal of the paper is to present a data-driven approach to label attribution. As such, the model description in Section~\ref{sec:model} is anchored on a dataset of users. Instead, we move to a slightly more abstract setting where we define a \textbf{history} $h$ as a nonempty sequence of elements of $\mathcal{X}$ that represents a sequence of displays. Recall that $\mathcal{X}$ denotes the set of potential displays. We let $\History$ be the set of all possible histories. Unlike the set of users, which we use in the main body of the paper and which is given by the data, we assume that the histories are drawn from some probability~$\proba$ that denotes the propensity of each history to appear in the historical logs. Additionally, each history $h \in \mathcal{H}$ is associated with an expected reward $R(h) \in \mathbb{R}^+$. One can map a user $u$ to a history $h$ and the corresponding reward $r_u$ to a particular realization of $R(h)$. However, different users with the same history might get different rewards and hence $R(h)$ captures an expected reward. For the rest of this section, we drop the users terminology.

\subsection{Notation and assumptions}

For any history $h$, we denote by $|h|$ its size and use the notation $h = [\boldsymbol{x}_1,\dots,\boldsymbol{x}_{|h|}]$, where $\boldsymbol{x}_i \in \mathcal{X}$ for $i \in \{1,\dots, |h|\}$, to denote its individual elements. Furthermore, we write $h \preccurlyeq h'$ if $h$ is a sub-history of $h'$, that is if $h$ is equal to the sequence of the $|h|$ first displays of $h'$.  

For any history $h$, we let $h_{[1,\dots, i]}$ be the projection of $h$ onto its first $i$ elements, i.e., for all $i \leq |h|$, we have $h_{[1,\dots,i]} = [\boldsymbol{x}_1,\dots,\boldsymbol{x}_i]$. With these notations, $h' \succcurlyeq h$ is equivalent to $|h'| \geq |h|$ and $h'_{[1,\dots,|h|]} = h$. Finally, for all $\boldsymbol{x} \in \mathcal{X}$ and $h \in \History$, let $h + [\boldsymbol{x}] = [\boldsymbol{x}_1,\dots,\boldsymbol{x}_{|h|},\boldsymbol{x}]$ denote the history obtained from adding a display $\boldsymbol{x}$ to a history $h$.
We end this section with two technical assumptions. 
\begin{assumption} \label{ass:finite}
$\mathcal{X}$ is a finite set.
\end{assumption}
Assumption~\ref{ass:finite} is not limiting in practice. Indeed, if some display features are continuous rather than categorical, they can be quantized~\citep{Chapelle2014b} in order to satisfy Assumption~\ref{ass:finite}. We also make the following assumption on the underlying probability distribution of histories.
\begin{assumption} \label{ass:proba}
If $\PP$ has support on a history, then it also has support on all its sub-histories. Namely, if $\;\PP(h) > 0$ for some $h \in \History$, then $\PP(h') > 0$ for all $h' \preccurlyeq h$.
 \end{assumption}
This is a mild assumption which is verified for various generative models used in the literature such as  Markovian models \citep{archak2010mining,anderl2016mapping,Besbes2019a} or  point process models \citep{xu2014path}.

\subsection{Valuation and perfect valuation}

Using the history notation, for every $(\boldsymbol{x},h) \in \mathcal{X} \times \mathcal{H}$, we let $V(\boldsymbol{x}|h)$ denotes the expected value of the display $\boldsymbol{x}$ given the history $h$. Note that in practice, making this dependence is not critical as $\boldsymbol{x}$ often include that information through features such as number of ads seen or number of ads clicked. However, we make this dependence explicit for the purpose of making our theoretical presentation more rigorous and precise. We also revisit the definition of label attribution as follows.
\begin{definition}[label attribution] \label{def:label_attribution_app}
A \textbf{label attribution} $\Mapping$ is a  mapping from  $\mathbb{N} \times \mathcal{H}$ to  $\RR^+$ that satisfies the following properties.
\begin{enumerate}
    \item $\forall h \in \mathcal{H}$, $\forall i > |h|$, $\Mapping(i|h) = 0$.
    \item $\forall h \in \mathcal{H}$, $R(h) = \sum_{i=1}^{|h|} \mu(i,h)$.
\end{enumerate}
\end{definition}
For a label attribution $\mu$ and history $h = [\boldsymbol{x}_1,\dots,\boldsymbol{x}_{|h|}]$, $\mu(i,h)$ represents the label of the $i^{th}$ display $\boldsymbol{x}_i$, i.e., the part of the reward $R(h)$ attributed to $\boldsymbol{x}_i$.

For a given observed history $h$ and display opportunity $\boldsymbol{x}$, recall that $H \succeq h+[\boldsymbol{x}]$ means that $H$ starts with the sequence of displays of $h+[\boldsymbol{x}]$, where $h+[\boldsymbol{x}]$ is the history obtained from adding a display $\boldsymbol{x}$ to a history $h$. We next define the ideal value learned by a perfect algorithm.

\begin{definition}[Associated valuation] \label{def:associated}
For every label attribution $\Mapping$ and probability distribution $\PP$, the associated valuation $\Valuation_{\PP}^{\Mapping}(x|h)$ is defined, for all $(x,h) \in \mathcal{X} \times \History$, as
\begin{equation} \label{eq:associated_valuation}
   \Valuation_{\PP}^{\Mapping}(\boldsymbol{x}|h) =  \EE_{H \sim \PP} \left[\Mapping(|h| + 1,H)\ \left| \ H \succcurlyeq h + [\boldsymbol{x}] \right. \right].
\end{equation}
\end{definition}
Importantly, note that $ \Valuation_{\PP}^{\Mapping}$ depends on the label attribution choice $\Mapping$ as well as the distribution $\PP$. Another way to interpret the definition of the associated  valuation is that when the bidder observes a display opportunity $\boldsymbol{x}$ after a history $\history$ has already happened, the bidder does not know whether additional opportunities will come in the future. If no other opportunity arises, then the display should be valued $\Mapping(|h|+1,h+[\boldsymbol{x}])$. On the other hand, if more displays are added, then the display should be valued $\Mapping(|h|+1,H)$ for some $H \succcurlyeq h+[\boldsymbol{x}]$. 

For instance, the valuation associated with the last touch label attribution is given by
\begin{align*}
    \Valuation^{\Mapping^{\sf LT}}_{\PP} (\boldsymbol{x}|h) = \Reward(h+[\boldsymbol{x}]) \cdot \PP (H = h + [\boldsymbol{x}]\; | \; H \succcurlyeq h + [\boldsymbol{x}]).
\end{align*}
If the label attribution is driven by the last touch philosophy, then an opportunity is evaluated through its probability of exactly preceding a conversion, which is precisely what the quantity $\PP (H = h + [\boldsymbol{x}]\; | \; H \succcurlyeq h + [\boldsymbol{x}])$ is capturing. It is worth noticing that this valuation depends  on the distribution of future displays. This is precisely the caveat observed in the Introduction as this dependence might lead the bidder to overvalue displays that come late in the conversion funnel.

\subsection{Additive valuation and fixed point characterization}
We can now properly define the additive valuation $V^{\sf FP}$ as follows. For all $(\boldsymbol{x},h) \in \mathcal{X} \times \mathcal{H}$,
\begin{align} \label{eq:additivity_app}
    V^{\sf FP}(\boldsymbol{x}|h) = R(h + [\boldsymbol{x}]) - R(h).
\end{align}
Note that in the paper, additivity is a property one can strive for. However, we cannot properly define an additive valuation per se since the set of users, or realized history, may not be complete. On the other hand, in this abstract setting where we have a reward function $R(h)$ for all history $h \in \mathcal{H}$, we can formally define the additive valuation $V^{\sf FP}$. In turn, we can define the additive label attribution $\mu^{\sf add}$ through a fixed point equation. In particular, for all $h = [\boldsymbol{x}_1,\dots,\boldsymbol{x}_{|h|}] \in \mathcal{H}$ and $i \leq |h|$, 
\begin{align} \label{eq:fixed_point}
    \Mapping^{\sf FP}(i,h) =  \frac{\Valuation^{\sf FP} \big(\boldsymbol{x}_i\;|\;h_{[1,\dots,i-1]}\big)}{\sum_{j=1}^{|h|}\Valuation^{\sf FP} \big(\boldsymbol{x}_j\;|\;h_{[1,\dots,j-1]}\big)} \cdot \Reward(h).
\end{align}
Note the obvious parallel with Equation~\eqref{eq:mtc_proof}. We are now ready to formalize the relation between additivity and proportional label attribution through the fixed point equation.
\begin{proposition} \label{prop:fixed_point}
$V^{\mu}_{\PP} = V^{\sf FP}  \iff \mu = \mu^{\sf FP}$
\end{proposition}
In words, Proposition~\ref{prop:fixed_point} states that a label attribution leads to the additive valuation, i.e., the additive valuation is associated to the additive label attribution, if and only if the label attribution and valuation satisfy the fixed point equation~\eqref{eq:fixed_point}.

\subsection{Convergence of the algorithm}

We now formally justify the design of our algorithm. In particular, consider the following iterative procedure:
\begin{align} \label{eq:iterative}
    \Valuation^{(k+1)}(\boldsymbol{x} | h) =\Valuation_{\PP}^{\Mapping^{(k)}} (\boldsymbol{x}|h), \ \forall (\boldsymbol{x},h) \in \mathcal{X} \times \History,
\end{align}
where for all $h= [\boldsymbol{x}_1,\dots,\boldsymbol{x}_{|h|}]$ and $i \leq |h|$,
\begin{align} \label{eq:iterative2}
    \Mapping^{(k)}(i,h) =  \frac{\Valuation^{(k)}(\boldsymbol{x}_i \ | \ h_{[1,\dots,i-1]})}{\sum_{j=1}^{|h|}\Valuation^{(k)}(\boldsymbol{x}_j \ | \ h_{[1,\dots,j-1]})} \cdot \Reward(h).
\end{align}
Recall that for any $\Mapping$ and $h \in \History$, the valuation $\Valuation^{\Mapping}_{\PP}(h)$, defined in Equation~\eqref{eq:associated_valuation}, is the expected reward, where the expectation is conditioned on $H\succcurlyeq h$. Note that Equation~\eqref{eq:iterative} corresponds to the prediction step of \textsf{FiPLA}  Algorithm in our ideal setting where one can learn perfectly. We also initialize the procedure with some label attribution $\Mapping^{(0)}$.
\begin{proposition} \label{prop:convergence}
For any $\Mapping^{(0)}$ such that $\Valuation_{\PP}^{\Mapping^{(0)}} (\boldsymbol{x}|h)>0$ for all $(\boldsymbol{x},h) \in \mathcal{X} \times \History$, the procedure defined in Equations~\eqref{eq:iterative}-\eqref{eq:iterative2} converges. Moreover, $\Mapping^{(k)} \to \Mapping^{\sf FP}$ when $k \to \infty$.
\end{proposition}
Note that the technical assumption $\Valuation_{\PP}^{\Mapping^{(0)}} (x|h)>0$ on the initial valuation is not required to prove convergence but only to prove that the iterates converge to the robust label attribution. Indeed, without this assumption, some coordinates can get stuck at their initial value. Proposition~\ref{prop:convergence} justifies the convergence of our proposed \textsf{FiPLA} Algorithm in this abstract setting.

\paragraph{Remark.} Consider a setting where we only have one sample with history $h=(\boldsymbol{x}_1, \dots, \boldsymbol{x}_I)$ but do not have data for any subhistory $h=(\boldsymbol{x}_1, \dots, \boldsymbol{x}_i)$, $\forall i < I$. In this case, Assumption~2 is does not hold. However, for practical purposes, having access to a single sequence of displays is one setting where we do not have enough data to know what to attribute. Moreover, in such case, the \textsf{FiPLA} Algorithm would be stuck at the initial guess for $\mu^0$. This means that the algorithm would revert to the heuristic used for initialization, such as any rule-based heuristic like last touch or uniform attribution. This looks like a reasonable fallback: when there is not enough data to learn anything, we do not learn anything beyond the initialization. 

\subsection{Additional properties}

We conclude this section, by showing that the additive valuation has several notable properties. We begin by proving that bidding according to the additive valuation is the only myopic optimal bidding strategy and then show that the additive valuation is the only valuation that does not depend on the probability distribution $\PP$ making it very robust from a learning perspective.

\subsubsection{Myopic optimality.} We first show that bidding according to the robust valuation is myopic optimal, i.e. optimal if the sequence of displays terminates without any future opportunities.
\begin{proposition} \label{prop:additivity}
$\Valuation$ satisfies the additivity property \textbf{if and only if} it is myopic optimal, i.e. if in a second-price setting, bidding $\Valuation$ is weakly dominant for any competition profile assuming there will be no other future opportunities.
\end{proposition}

We acknowledge that this notion of optimality is -- as the name suggests -- rather narrow: the potential combined effects of the decision at stake (the display to be bought) and the future ones are neglected. Nevertheless, we believe this notion has some merit. In particular, the standard notion of optimality implies that the best decision at a given time depends on what will be done in the future.  However, solving the complete optimization problem is in practice often intractable and therefore, one can only hope for approximate solutions. By contrast, as we will discuss next, our proposed strategy does not depend on the probability distribution $\PP$ (and hence, on the impact of our bidding strategy in the future), making it more robust. Finally, our aim is also to shift the discussion to principled approaches to label attribution, very much in the same spirit than \cite{Besbes2019a}. Here, we prove that there is a unique valuation which is myopic optimal.

\subsubsection{Distributional robustness.} We next show that the robust valuation does not depend on the underlying probability distribution $\PP$, i.e., the propensity of each history to appear in the historical logs.

\begin{proposition} \label{prop:robust}
$\Valuation$ satisfies the additivity property \textbf{if and only if} $\Valuation$ is an associated valuation which does not depend on $\PP$ but only on the reward function $\Reward$, i.e., there exists a label attribution $\Mapping$ such that $\Valuation = \Valuation^{\Mapping}_{\PP}  =  \Valuation^{\Mapping}_{\QQ} $ for any other probability distribution $\QQ$ over $\History$ that has the same support as $\PP$.
\end{proposition}
Surprisingly, the additivity property also leads to distributional robustness. From a learning perspective, this means that this approach alleviates the need of learning the underlying probability distribution $\PP$. 
In particular, if the environment changes and we receive a new probability distribution on the histories $\QQ$, then the expected reward can still be computed using the formula
$$
    \EE_{H\sim \QQ} \ \Reward(H) = \EE_{H \sim \QQ} \ \sum_{i=1}^{|H|} \Valuation^{\Mapping}_{\PP}(H_i \ | \ H_{[1,\dots,i-1]}),$$
even when $\PP\neq\QQ$. We emphasize that this is not true for any other valuation mechanism. Finally, distributional invariance is often related to causality~\citep{peters2016causal, arjovsky2019invariant}, and it would be an interesting research direction to further explore these connections.

\section{Proofs} \label{sec:proof}

\subsection{Proof of Proposition~\ref{prop:fixed_point}} \label{app:fixed_point}

We first show that the valuation associated to the additive label attribution is the additive valuation. First, by inspecting Equation~\eqref{eq:fixed_point}, it is clear that $\Mapping^{\sf add}$ satisfies Definition~\ref{def:label_attribution_app}. Now, for all $(\boldsymbol{x},h) \in \mathcal{X} \times \mathcal{H}$, we have, using Equation~\eqref{eq:associated_valuation},
\begin{align*}
  \Valuation_{\PP}^{\Mapping^{\sf FP}}(\boldsymbol{x}\;|\;h )  
  &= 
  \EE_{H \sim \PP} \left[ \left. \frac{\Valuation^{\sf FP}(\boldsymbol{x} | H_{[1,\dots,|h|]})}{\sum_{j=1}^{|H|} \Valuation^{\sf FP}(H_j | H_{[1,\dots,j-1]})} \cdot \Reward(H) \ \right| \  H \succcurlyeq  h  + [\boldsymbol{x}] \right]  \\
  &= \Valuation^{\sf FP}(\boldsymbol{x} | h) \cdot \EE_{H \sim \PP} \left[\left. \frac{\Reward(H)}{\sum_{j=1}^{|H|} \Valuation^{\sf FP}(H_j | H_{[1,\dots,j-1]})} \right| H \succcurlyeq h \right] \\
  &= \Valuation^{\sf FP}(\boldsymbol{x} | h) ,
\end{align*}
where the last equality follows Equation~\ref{eq:additivity_app}.
  
We next show the reverse implication. We assume we have a label attribution $\mu$ and valuation $V$ such that the fixed point equation is verified, i.e., for all $h= [\boldsymbol{x}_1,\dots,\boldsymbol{x}_{|h|}]$ and $i \leq |h|$,
  \[
\Mapping(i,h) =  \frac{\Valuation\big(h_i\;|\;h_{[1,\dots,i-1]}\big)}{\sum_{j=1}^{|h|}\Valuation\big(\boldsymbol{x}_j\;|\;h_{[1,\dots,j-1]}\big)} \cdot \Reward(h).
\]
Using Definition~\ref{def:associated}, we have for all $(\boldsymbol{x},h) \in \mathcal{X} \times \History$,
\[
\Valuation_{\PP}^{\Mapping}(\boldsymbol{x}|h) =  \EE_{H \sim \PP} \left[\Mapping(|h| + 1,H)\ \left| \ H \succcurlyeq h + [\boldsymbol{x}] \right. \right].
\]
Consequently, for all for all $h= [\boldsymbol{x}_1,\dots,\boldsymbol{x}_{|h|}]$ and $i \leq |h|$, we obtain
\begin{align*}
\sum_{j=1}^{|h|}\Valuation\big(\boldsymbol{x}_j\;|\;h_{[1,\dots,j-1]}\big) 
&= \sum_{j=1}^{|h| }\EE_{H \sim \PP} \left[\Mapping(j,H)\ \left| \ H \succcurlyeq h_{[1, \dots, j]} \right. \right] \\
&= \sum_{i=1}^{|h| }\EE_{H \sim \PP} \left[\frac{\Valuation\big(\boldsymbol{x}_i\;|\;h_{[1,\dots,i-1]}\big)}{\sum_{j=1}^{|h|}\Valuation\big(\boldsymbol{x}_j\;|\;h_{[1,\dots,j-1]}\big)} \cdot \Reward(h) \left| \ H \succcurlyeq h_{[1, \dots, i]} \right. \right]\\
&= R(h),
\end{align*}
where we obtain the last equality by interchanging the expectation and sum operators. This concludes the proof.

\subsection{Proof of Proposition~\ref{prop:convergence}}
\label{sec:proof-pop-convergence}

To prove the convergence of the iterative procedure defined in Equations~\eqref{eq:iterative}-\eqref{eq:iterative2}, we show that it corresponds to the iterates of a majorize-minorize (MM) algorithm. We remind the reader of the MM philosophy and refer to \cite{hunter2004tutorial} for more details. An MM algorithm aims at finding the maximizer $\theta^*$ of a function $f(\theta)$. For that purpose, a surrogate function $g(\theta|\hat{\theta})$ that depends on a current estimate $\hat{\theta}$ and is typically easier to maximize is identified and maximized instead. The key is to choose a surrogate function that minorizes the objective function.  A typical iteration of the MM algorithm can be written as follows. Given an estimate $\theta^{(m)}$, we let 
\begin{align*}
    \theta^{(m+1)} = \argmax \limits_{\theta} g(\theta | \theta^{(m)}).
\end{align*}
The procedure is guaranteed to converge to a local maximum of $f$ as long as the following are satisfied:
\begin{enumerate}
    \item $f(\theta)$ is concave,
    \item $g(\theta^{(m)}|\theta^{(m)}) = f(\theta^{(m)})$,
    \item $g(\theta|\theta^{(m)}) \leq f(\theta), \forall \theta$.
\end{enumerate}
We show that our iterative procedure is a special case of an MM algorithm by letting for all $\Valuation : \mathcal{X} \times \History \mapsto \mathbb{R}^+$,
\begin{align*}
f(\Valuation) =&\EE_{H\sim\PP} \left[ \Reward(H) \cdot \ln \left(\sum_{q \in \{1,\dots,|H|\} }\Valuation \left( H_q  |  H_{[1,\dots,q-1]} \right) \right)  - \sum_{q \in \{1,\dots,|H|\} } \Valuation \left( H_q  |  H_{[1,\dots,q-1]}\right)\right], \\
g(\Valuation |\hat{\Valuation}) =& \EE_{H\sim\PP} \Bigg[ \sum_{j\in \{1,\dots,|H|\}} \Bigg\{ \frac{\hat{\Valuation}\left(H_j |  H_{[1,\dots,j-1]} \right) \cdot \Reward(H)}{\sum_{q\in \{1,\dots,|H|\}} \hat{\Valuation}\left(H_q |  H_{[1,\dots,q-1]} \right)} \cdot \ln \left(\frac{\Valuation(H_j  |  H_{[1,\dots,j-1]})}{\hat{\Valuation}(H_j |  H_{[1,\dots,j-1]})}\sum_{q\in \{1,\dots,|H|\}}\hat{\Valuation}(H_q | H_{[1,\dots,q-1]}) \right) \\
& \ \ \ \  - \Valuation(H_j|H_{[1,\dots,j-1]}) \Bigg\} \Bigg].
\end{align*}
Since the space of histories is finite, $\Valuation$ can always be mapped to a finite dimension vector. We do however keep the function notation for simplicity. First, note that $f$ is strictly concave as a linear combination of concave functions.  Second, observe that for all $\Valuation$,
\begin{align*}
    g(\Valuation|\Valuation) =& \EE_{H\sim\PP} \Bigg[ \sum_{j\in \{1,\dots,|H|\}}  \Bigg\{  \frac{\Valuation(H_j|H_{[1,\dots,j-1]}) \cdot \Reward(H)}{\sum_{q\in \{1,\dots,|H|\}} \Valuation(H_q | H_{[1,\dots,q-1]})} \cdot \ln \left(\frac{\Valuation(H_j|H_{[1,\dots,j-1]})}{\Valuation(H_j|H_{[1,\dots,j-1]})}\sum_{q\in \{1,\dots,|H|\} } \Valuation(H_q | H_{[1,\dots,q-1]}) \right)  \\
    & \ \ \ \ - \Valuation(H_j|H_{[1,\dots,j-1]}) \Bigg\} \Bigg] \\
    =&\EE_{H\sim\PP} \Bigg[ \Reward(H) \cdot \ln \Bigg(\sum_{q\in \{1,\dots,|H|\} } \Valuation(H_q|H_{[1,\dots,q-1]}) \Bigg) \cdot  \sum_{j\in \{1,\dots,|H|\}}  \frac{\Valuation(H_j|H_{[1,\dots,j-1]}) }{\sum_{q\in \{1,\dots,|H|\}} \Valuation(H_q | H_{[1,\dots,q-1]})}    \\
    &\ \ \ \ - \sum_{j\in \{1,\dots,|H|\}} \Valuation(H_j|H_{[1,\dots,j-1]}) \Bigg]\\
    =&f(\Valuation).
\end{align*}

Third, the inequality $f(\Valuation) \geq g(\Valuation|\hat{\Valuation})$ follows by using the concavity of the logarithm and Jensen inequality on the random variable $X$ which take value $$\dfrac{\Valuation(H_j|H_{[1,\dots,j-1]})}{\hat{\Valuation}(H_j|H_{[1,\dots,j-1]})} \cdot \sum_{q\in \{1,\dots,|H|\} }\hat{\Valuation}(H_q|H_{[1,\dots,q-1]})$$ with probability $\hat{\Valuation}(H_j|H_{[1,\dots,j-1]})/ \sum_{q\in \{1,\dots,|H|\}} \hat{\Valuation}(H_q|H_{[1,\dots,q-1]}) $. Finally, for any $\hat{\Valuation}$, we have   $\mbox{argmax}_{\Valuation} g(\Valuation|\hat{\Valuation}) = \Valuation_{\PP}^{\Mapping_{\hat{\Valuation}}}$ by Lemma \ref{lemma:gOptim} below, where for all $\hat{\Valuation} : \History \mapsto \RR$, we define $\mu_{\hat{V}}$ as follows. For all $h = [\boldsymbol{x}_1,\dots,\boldsymbol{x}_{|h|}] \in \mathcal{H}$ and $i \leq |h|$, 
\begin{align} \label{eq:fixed_point}
    \mu_{\hat{V}} (i,h) =  \frac{\hat{V} \big(\boldsymbol{x}_i\;|\;h_{[1,\dots,i-1]}\big)}{\sum_{j=1}^{|h|}\hat{V} \big(\boldsymbol{x}_j\;|\;h_{[1,\dots,j-1]}\big)} \cdot \Reward(h).
\end{align}
Using \cite{hunter2004tutorial}, we conclude that the sequence of $\Valuation^{(k)}$ corresponds to the iterates of an MM algorithm with strictly concave criteria, and hence converges. This concludes the proof of Proposition~\ref{prop:convergence}.

\begin{lemma}
\label{lemma:gOptim}
For any $\hat{\Valuation} : \History \mapsto \RR$, $\argmax_{\Valuation} g(\Valuation | \hat{\Valuation}) =\Valuation_{\PP}^{\Mapping_{\hat{\Valuation}}}$.
\end{lemma}
\proof{Proof.}
Since $g(\Valuation| \hat{\Valuation})$ is concave in $\Valuation$, it suffices to show that $\Valuation_{\PP}^{\Mapping_{\hat{\Valuation}}}$ satisfies the first order condition. In particular, for any $(\boldsymbol{x},h) \in \mathcal{X} \times \History$,
\begin{align*}
    \frac{\partial g(\Valuation| \hat{\Valuation})}{\partial \Valuation(\boldsymbol{x}|h)} =& \frac{\partial }{\partial \Valuation(\boldsymbol{x}|h)} \Bigg( \sum \limits_{t}  \PP (t) \cdot \sum_{j\in \{1,\dots,|t|\}} \Bigg\{ \frac{\hat{\Valuation}(t_j|t_{[1,\dots,j-1]}) \cdot \Reward(t)}{\sum_{q\in \{1,\dots,|t|\}} \hat{\Valuation}(t_q| t_{[1,\dots,q-1]})} \cdot \ln \left(\frac{\Valuation(t_j|t_{[1,\dots,j-1]})}{\hat{\Valuation}(t_j|t_{[1,\dots,j-1]})}\sum_{q\in \{1,\dots,|t|\} }\hat{\Valuation}(t_q|t_{[1,\dots,q-1]}) \right)  \\
    & \ \ \ \ - \Valuation(t_j|t_{[1,\dots,j-1]}) \Bigg\} \Bigg)
\end{align*}
Note that $h$ appears in the inner sum if and only if $t \succcurlyeq h+[\boldsymbol{x}]$. Therefore, we can write
\begin{align*}
    \frac{\partial g(\Valuation| \hat{\Valuation})}{\partial \Valuation(\boldsymbol{x}|h)} =& \frac{\partial }{\partial \Valuation(\boldsymbol{x}|h)} \Bigg( \sum \limits_{t \succcurlyeq h+[\boldsymbol{x}]}  \PP (t) \cdot \Bigg\{ \frac{\hat{\Valuation}(\boldsymbol{x}|h) \cdot \Reward(t)}{\sum_{q\in \{1,\dots,|t|\}} \hat{\Valuation}(t_q|t_{[1,\dots,q-1]})} \cdot \ln \left(\frac{\Valuation(\boldsymbol{x}|h)}{\hat{\Valuation}(\boldsymbol{x}|h)}\sum_{q\in \{1,\dots,|t|\} }\hat{\Valuation}(t_q|t_{[1,\dots,q-1]}) \right) \\ & \ \ \ \ - \Valuation(\boldsymbol{x}|h) \Bigg\} \Bigg).
\end{align*}
Exchanging the derivative and sum operators yields
\begin{align*}
    \frac{\partial g(\Valuation| \hat{\Valuation})}{\partial \Valuation(\boldsymbol{x}|h)} =  \sum \limits_{t \succcurlyeq h+[\boldsymbol{x}]}  \PP (t) \cdot \Bigg\{ \frac{\hat{\Valuation}(\boldsymbol{x}|h) \cdot \Reward(t)}{\sum_{q\in \{1,\dots,|t|\}} \hat{\Valuation}(t_q|t_{[1,\dots,q-1]})} \cdot \dfrac{1}{\Valuation(\boldsymbol{x}|h)}  - 1 \Bigg\}.
\end{align*}
Setting the above to zero for the first order condition then implies that for all $(\boldsymbol{x},h) \in \mathcal{X} \times \History$, we have
\begin{align*}
    \Valuation(\boldsymbol{x}|h) &= \dfrac{1}{\sum \limits_{t \succcurlyeq h+[\boldsymbol{x}]}  \PP (t)} \cdot \left( \sum \limits_{t \succcurlyeq h+[\boldsymbol{x}]}  \PP (t) \cdot  \frac{\hat{\Valuation}(\boldsymbol{x}|h) \cdot \Reward(t)}{\sum_{q\in \{1,\dots,|t|\}} \hat{\Valuation}(t_q|t_{[1,\dots,q-1]})}\right) \\
    &= \EE_{H\sim\PP} \left[ \frac{\hat{\Valuation}(\boldsymbol{x}|h)}{\sum_{q\in \{1,\dots|H|\}} \hat{\Valuation}(H_q|H_{[1,\dots,q-1]})} \cdot \Reward(H) \; \big| \; H \succcurlyeq h +[\boldsymbol{x}] \right] \\
    &= \Valuation_{\PP}^{\Mapping_{\hat{\Valuation}}}(\boldsymbol{x}|h) .
\end{align*} 
This concludes the proof. \halmos
\endproof

\subsection{Proof of Proposition~\ref{prop:additivity}} \label{app:additivity}

Fix a history $h \in \History$ and display $\boldsymbol{x} \in \mathcal{X}$. Recall that bidding $\Valuation(\boldsymbol{x}|h)$ is myopic weakly dominant if and only if, in a second price setting, bidding $\Valuation(\boldsymbol{x}|h)$ is weakly dominant optimal for any competition profile. For a given competition profile, assume that the highest bid $t$ among competition follows a density $g$. The payoff of the auction is given by $(\Reward(h+[\boldsymbol{x}])-t)$ if the auction is won, i.e., if the bid $\Valuation(\boldsymbol{x}|h)$ exceeds $t$, and $\Reward(h)$ otherwise. The optimal bid $\Valuation(\boldsymbol{x}|h)$ must therefore maximize the following 
        \begin{equation*}
                    \int_0^{\Valuation(\boldsymbol{x}|h)}  (\Reward(h+[\boldsymbol{x}])-t)\cdot  g(t)\cdot dt  + \int_{\Valuation(\boldsymbol{x}|h)}^{+\infty} \Reward(h) \cdot  g(t) \cdot d t ,
        \end{equation*}
        where the first integral corresponds to the set of competition prices where we win the auction and the second term correspond to competition bids above $\Valuation$. In the latter case, we obtain $\Reward(h)$ because of the missed opportunity.  Writing the first order condition yields 
                \begin{equation*}
                    (\Reward(h+[\boldsymbol{x}])-\Valuation(\boldsymbol{x}|h) - \Reward(h)) \cdot g(\Valuation(\boldsymbol{x}|h)) = 0
        \end{equation*}
        for any $g$, hence, bidding
          $\Valuation(\boldsymbol{x}|h)$ is myopic weakly dominant if and only if  $\Reward(h+[\boldsymbol{x}])-\Valuation(h) - \Reward(h) = 0$.

\subsection{Proof of Proposition~\ref{prop:robust}} \label{app:robust}

By Proposition \ref{prop:fixed_point}, if $\Valuation$ is myopic optimal, then it is associated with the additive  attribution given in Equation~\eqref{eq:additivity_app}, which does not depend on $\PP$. We therefore only have to prove the reverse implication which we do next.

Assume that $\Valuation$ does not depend on $\PP$ but only on the reward function $\Reward$, i.e., there exists an internal attribution $\Mapping$ such that $\Valuation = \Valuation^{\Mapping}_{\PP}  =  \Valuation^{\Mapping}_{\QQ} $ for any other probability distribution $\QQ$ over $\History$. We begin by showing that there exists a unique valuation which is distributionally robust, i.e., if we let $\Mapping$ its associated  attribution, we must have for any probability $\PP$ and $h \in \History$
\begin{align*}
    \Valuation (h) =  \Valuation^{\Mapping}_{\PP} (h) = \EE_{H \sim \PP} [ \Mapping(|h|,H)|H \succcurlyeq h].
\end{align*}
The associated attribution defined by Equation~\eqref{eq:additivity_app} is distributionally robust since the expression of $\Mapping$ does not depend on $\PP$ which shows existence. We show unicity by contradiction. Assume that there are two such internal attributions $\Mapping^1$ and $\Mapping^2$. For all $\PP$, $h$ and $j\leq |h|$, we must have 
\begin{align*}
    \Valuation(h_j | h_{[1,\dots,j-1]}) = \EE_{H \sim \PP} [ \Mapping^1(j,H)|H \succcurlyeq h_{[1,\dots,j]}] = \EE_{H \sim \PP} [ \Mapping^2(j,H)|H \succcurlyeq h_{[1,\dots,j]}].
\end{align*}
Consider a probability $\QQ$ such that the weight of all elements $H\succcurlyeq h_{[1,\dots,j]}$ is moved to $h_{[1, \dots, j]}$. Then, we have for all $h$ and $j \leq |h|$, $\Mapping^1(j,h) = \Mapping^2(j,h)$. Note that since the probability $\QQ$ must satisfy Assumption \ref{ass:proba}, a limiting argument yields the result.

\end{APPENDICES}

\end{document}